\documentclass[12pt,preprint]{aastex}

\doublespace

\def\be{\begin{equation}}
\def\ee{\end{equation}}
\def\etal{~{\it et al.\ }}
\def\bea{\begin{eqnarray}}
\def\eea{\end{eqnarray}}
\def\noi{\noindent}

\shorttitle{Eccentric Planetary Rings}
\shortauthors{J.C.B. Papaloizou and  M.D. Melita}

\begin{document}

\centerline{\bf{\Large Structuring eccentric-narrow planetary rings}}
\vspace{0.5cm}
\centerline{{J.C.B. Papaloizou and  M.D. Melita}}
\vspace{0.5cm}
\centerline{\it{Astronomy Unit.}}
\centerline{\it School of Mathematical Sciences.}
\centerline{\it Queen Mary. University of London.}
\centerline{\it Mile End Rd.}
\centerline{\it {London E1 4NS. U.K.}}
\vspace{140mm}
\centerline{49 pages, 1 Figure \hspace{100mm}}
%
%\email{M.D.Melita@qmul.ac.uk}
%\centerline {\bf{jcbp@maths.qmul.ac.uk}}
\newpage
\centerline{ {\bf Proposed Running Head:} 
{\it Structuring eccentric-narrow planetary rings }\hspace{30mm}}
\vspace{30mm}
\centerline {\bf{Editorial correspondence to:}\hspace{100mm}}
\vspace{0.2cm}
\centerline  {J.C.B. Papaloizou \hspace{100mm}}
\vspace{0.2cm}
\centerline{\it{Astronomy Unit.}\hspace{100mm}}
\centerline{\it School of Mathematical Sciences.\hspace{100mm}}
\centerline{\it Queen Mary. University of London.\hspace{100mm}}
\centerline{\it Mile End Rd.\hspace{100mm}}
\centerline{\it {London E1 4NS. U.K.}\hspace{100mm}}
%
%\email{M.D.Melita@qmul.ac.uk}
\vspace{3mm}
\centerline {Telephone {\bf +44 (0)20 7882 5446}\hspace{100mm}}
\centerline{  Fax \hspace{10mm} {\bf +44 (0)20 8983 3522}\hspace{100mm}}
\centerline {email:  \bf{jcbp@maths.qmul.ac.uk}\hspace{100mm}}

\begin{abstract}
A simple and general description  of the dynamics of a narrow-eccentric ring is
presented. We view an eccentric ring 
which precesses uniformly at a
slow rate as exhibiting 
a global $m=1$ mode, which can be seen as originating from a standing
wave superposed on an axisymmetric background.  We  adopt a continuum description
using the language of fluid dynamics which gives equivalent results for the
secular dynamics of thin rings  as the
the well known description in terms of 
a set of discrete elliptical streamlines formulated by Goldreich and Tremaine (1979). 
We use this to discuss the non linear mode interactions that  appear
in the ring  through  the excitation of higher $m$ modes
because of the coupling of the $m=1$ mode with an external satellite potential, 
showing  that they   can lead to the excitation
of the $m=1$ mode through a feedback process.
In addition to the external perturbations by neighboring
satellites, our model includes effects due to inelastic inter-particle collisions.
Two main conditions for the ring to be able to maintain a 
steady $m=1$ normal mode
are obtained. One
can be expressed as an integral condition for the normal
mode pattern
to precess uniformly, which requires the correct balance between the
differential precession induced by the oblateness of the central planet,
self-gravity and collisional effects is the continuum form
of that obtained from the $N$ streamline model of Goldreich and Tremaine (1979).
The other condition,
not before examined in detail, is for the steady
maintenance 
of the non-zero radial action that the ring   contains because of its finite
normal mode.  This requires a balance   between injection due to
eccentric resonances arising from external satellites and additional
collisional damping associated with the presence of the $m=1$ mode.  We
estimate that such a balance  can occur 
in the $\epsilon-$ring of Uranus,
given its currently observed physical and orbital parameters.

\end{abstract}

\newpage

\section{Introduction}

The  nature of the dynamical mechanism that maintains the apse alignment of
narrow-eccentric planetary rings is one of the most interesting and
challenging problems of Celestial Mechanics. 
 
According to  the leading model (Goldreich and Tremaine 1979) the
self-gravity of the ring counter-acts the differential precession induced by
the oblateness of the central planet. Using this hypothesis, a prediction of
the total mass of the ring can be made, which, in general, is not in good
agreement with the inferred mass of the observed eccentric rings in the
Uranus system (Tyler\etal 1986, Graps\etal 1995, Goldreich and Porco 1987,
Porco and Goldreich 1987).
This led to the consideration of other factors that  might 
play an important role
in the dynamics.
In particular, at their narrowest point, the ring particles
are `close-packed'.  In such a situation particle interaction 
or pressure effects may affect the precession 
of particle orbits. A simple model where the {\it pinch} locks the
differential precession, was introduced by Dermott and Murray (1980). A more
global picture, including the effect of  stresses due to particle 
interactions and
neighboring satellite perturbations, which offered a better agreement with
the observations, has been produced by Borderies\etal (1983). Their
dynamical model is described in terms of mutually interacting {\it streamlines}
and the satellite interactions (see Goldreich  and Tremaine 1981)
are computed using a resonance-continuum
approximation. The standard self-gravity model was later revisited  by
Chiang and Goldreich (2000), who  considered the effects of collisions
near the edges, proposing that a sharp increase of an order of
magnitude in the surface density should be observed within the last few hundred
meters of the ring  edges. More 
recently, employing a pressure term that describes 
close-packing, Mosqueira and Estrada (2002) obtained surface-density
solutions that agree well with the currently available mass estimates.

However, several questions remain, such as
how
steady global $m=1$ modes are maintained by intercations with a satellite 
and mode couplings against dissipative processes.
In particular if such modes are associated with  particle close packing
at certain orbital phases, this is likely to produce enhanced dissipation
that has to be made up through the action of satellite torques.
Here we consider the issue
as to how the torque rate input
available from satellites can counteract such collisional dissipation. 

In this work we build, from first principles, a simple general 
continuum or fluid like model of
a narrow-eccentric ring. The eccentric pattern in the ring can be described
as being generated 
by a normal mode of oscillation of wave-number $m=1$ which may
be  considered to be a standing
wave. Dissipation can be allowed to occur
due to inter-particle collisions leading to a viscosity 
 which would lead to damping of the mode. 
However, this global $m=1$ mode can  also be perturbed by 
neighboring-shepherd satellites which can inject energy and angular
momentum through resonances. In this way losses due to particle collisions
may be balanced. It is that process that is the focus of this 
paper. Other possible mechanisms, such as  mode excitation through
self-excitation
through viscous overstability, that could arise
with  an appropriate dependence of viscosity on physical
state variables (see Papaloizou and Lin 1988, Longaretti \& Rappaport 1995),
are beyond the scope of this paper and accordingly  not investigated here.

To describe the ring perturbations
and   the $m=1$ mode we use
 the Lagrangian-displacement of the particle orbits
from their unperturbed circular  ones (for a similar treatment applied to
very diverse problems see for example Lebovitz 1967, Lynden-Bell and
Ostriker 1967, Friedman and Schutz 1978, Shu\etal 1985).

In section~\ref{LD} we set up the equations for the Lagrangian variations
starting from the  equations of motion in a 2D flat
disk approximation. We also compute the Lagrangian
variation of the satellite potential as seen by 
a particle as a consequence of the existence of the $m=1$ 
mode, and we discuss what non linear couplings appear
in the ring as a result including the excitation
of the $m=1$ mode through a feedback process.  In section~\ref{m1m} we derive 
the radial equation of motion for the  2D Lagrangian displacement
under the assumption that the ring is primarily in a $m=1$ normal mode. The
definition of the radial action in terms of 
the Lagrangian displacement
for small eccentricities is given in section~\ref{RA}, and its
rate of change is obtained.  We can then determine
a condition for the steady  maintenance of the
amplitude  or 
eccentricity  associated with 
the $m=1$ mode, which requires the external satellite input
to balance the dissipative effects due to collisions.  

\noindent In this paper, the application we consider is to
the $\epsilon$ ring of Uranus. As there are no effective  corotation resonances
in this ring if the eccentricity of the perturbing satellites is neglected
as is done here,  we shall consider only Lindblad
resonances and  defer consideration of corotation resonances
to future work.

The satellite torque
is obtained in section~\ref{SATT}. In section~\ref{RAST} we show that 
for the narrow rings considered here, the
satellite contribution to the rate of change of the radial action is just a
fraction of the corresponding satellite torque dependent
only on the relevant azimuthal mode number, $m.$ 

The additional condition for
the existence of the normal mode, i.e. the condition of uniform-precession,
is derived in section~\ref{UP}. The self-gravity term appearing in this
condition is computed in section~\ref{SGterm}. In section~\ref{vq} we show that
the eccentricity gradient is necessarily positive in a narrow-ring which 
in which  uniform precession is mainly maintained
 by self-gravity and we estimate its value in the linear
regime. 

In section~\ref{discu} we discuss our results and 
by  considering
the total ring radial action using a very simple N-body
approach, we 
illustrate the
global functioning of a narrow-eccentric ring.
We are able to obtain  both ring spreading and
the balance of satellite torques and  collisional dissipation
required to maintain ring eccentricity
in various
limiting cases of ring  evolution. Finally, we apply our
results to the $\epsilon-$ring of Uranus estimating 
that the balance between the satellite torque input and the dissipative
effects necessary to maintain its eccentricity  
can be established in this case.

%\section{Relation between rate of pumping of eccentric 
%	ring radial action
%	and satellite torques at eccentric Lindblad resonances} \label{ACT}
%

\section{Equations of Motion and Lagrangian Displacement}
\label{LD}
We adopt a continuum or fluid description of the system
from a Lagrangian viewpoint.
There is an issue about whether such a description is applicable to   planetary rings 
for which the particle collision time is typically comparable
to the orbital time. An approach based on taking moments
of the Boltzmann equation
(see Stewart, Lin  and Bodenheimer 1984 for a discussion)
yields fluid equations with a stress tensor that is
determined by the details of the  collisional
behavior and therefore uncertain.
However, this  consideration 
only affects viscous and pressure effects.
Phenomena such as density waves which control global modes
are independent of it (see Shu 1984 for more discussion).
Furthermore some viscous  phenomena like those considered
here are related to conservation laws so that the way they
enter is clear even if details are uncertain.
A Lagrangian description is the natural one because the
description of Keplerian orbits is straightforward and in a ring
for which particle motion deviates slightly from Keplerian,
torques due to satellites acting on the ring  etc. arise from changes as
seen by a moving fluid element.

Goldreich  and Tremaine (1981) and Borderies, Goldreich  and Tremaine (1983)
adopt an approach in which ring particles are assumed
to be on elliptical streamlines  with slowly
varying (compared to orbital times) osculating Keplerian elements.
In many ways that approach and the one followed here are similar.

\noindent But the assumption of slow variation means that disturbances
producing satellite torques are not included directly 
as they need to be for mode coupling. Also this model
is a discrete one consisting of $N$ streamlines.
This necessitates care with regard to the singular integrals,
that need to be taken in the principal value sense, that occur
when dealing with self-gravity (see section \ref{SGterm} below ).

We start from the basic equations of motion for a particle in Lagrangian
form in 2D:
\be 
{d^2 r\over dt^2} - r\left( {d\theta \over dt}\right)^2 = 
F_r -{\partial \psi \over \partial r}
\label{Mr}
\ee
\be 
r\ {d^2 \theta \over dt^2} + 2\left({d r \over dt}\right)\left( {d\theta
\over dt}\right) = 
F_{\theta} - {1\over r} {\partial \psi \over \partial \theta } 
\label{MTH}
\ee
\noi Here $(r,\theta)$ define the cylindrical coordinates of the
particle referred to an origin at the center of mass of the planet. Here
$\psi(r)$ denotes the gravitational potential due to both the central
planet, the neighboring satellites and the ring. In addition $(F_r,
F_{\theta})$ denote the radial and azimuthal components of any additional
force ${\bf F}$ per unit mass respectively.  This may arise through internal
interactions between particles that might lead to an effective pressure
and/or viscosity. However, we do not need to introduce such concepts in 
order to derive our condition determining the growth or decay of global
$m=1$ modes.

We introduce a Lagrangian description in which the system is supposed to be
perturbed from an axisymmetric state in which particles are in circular
motion with coordinates such that $r = r_0 ,$ $ \theta = \theta_0=
\Omega(r_0)t + \beta_0.$ Here $r_0$ is the fixed radius of the particle
concerned, $\Omega(r_0)$ is the angular velocity and $\beta_0$ is a phase
factor labeling each particle. In keeping with a Lagrangian description
$(r_0, \beta_0 )$ are conserved quantities for a particular particle and so
may be used to label it.

In order to describe the system when it is perturbed from the axisymmetric
state we introduce the components of the Lagrangian displacement $
{\mbox{\boldmath$\xi$}} = (\xi_r, \xi_{\theta}).$ These are such that the
coordinates of each particle satisfy:
\be r = r_0 + \xi_r,\ee and \be r_0(\theta -\theta_0 ) = \xi_{\theta}.\ee

To obtain equations for $\xi_r$ and $\xi_{\theta}$ we take variations of
Eq.'s~({\ref{Mr}) and (\ref{MTH}). We do this by applying the Lagrangian
difference operator, $\Delta$, as defined by Lebovitz (1961) to both sides
of Eq.'s~(\ref{Mr}) and~(\ref{MTH}). For a given quantity $Q$, the
variation $\Delta(Q)$ is defined by: \be \Delta(Q) = Q\left( {\bf r}_0 + {\bf
\xi} \right) - Q_0\left( {\bf r}_0 \right), \label{Lagrdef} \ee where $Q$
and $Q_0$ are the values of the given physical quantity in the perturbed and
unperturbed flow respectively.  In contrast, the Eulerian difference
operator is defined as: \be \delta(Q) = Q\left( {\bf r}_0 \right) - Q_0\left(
{\bf r}_0 \right). \ee Thus, to first order in 
${\mbox{\boldmath$\xi$}}$  they are related through: \be
\Delta =\delta + {\mbox{\boldmath$\xi$}} \ .\ \nabla \ee which gives the
linear form of the Lagrangian difference operator.

\subsection {Equations for the Lagrangian displacement} Following Shu\etal
(1985) we assume that the components of the displacement are small enough
that they can be treated as linear in the sense that
$|{\mbox{\boldmath$\xi$}}/r_0| << 1.$ On the other hand the radial gradient
of the radial displacement may be large so that $|(\partial \xi_r/\partial
r_0)|$ may be of order unity. The significance of these assumptions is that
although the ring eccentricity is assumed to be everywhere small, the ring
surface density perturbation induced by it may be of order unity. Adopting
them enables us to perform the variation in the accelerations using the
linear form of the difference operator as described above, wherever radial
gradients are not involved. These then satisfy: \be {d^2 \xi_r\over dt^2}
-2\Omega {d \xi_{\theta} \over dt}+ 2\xi_r r_0 \Omega {d \Omega \over dr_0}
= f_{r} - \Delta \left({\partial \psi' \over \partial r} \right)
\label{pr}\ee \be {d^2 \xi_\theta \over dt^2} + 2\Omega {d \xi_r \over dt} =
f_{\theta} - \Delta \left( {1\over r} {\partial \psi' \over \partial \theta
} \right). \label{pth0} \ee 

Here the potential due to the satellites, $\psi_{s}$, and
that due to the self-gravity of the ring, $\psi_{SG}$, are included in
$\psi'.$ Thus $\psi' = \psi_{SG} + \psi{s}.$ The quantities $f_r =
\Delta(F_{r}),\ f_\theta = \Delta(F_{\theta})$ denote the variational
components of the force per unit 
mass due to particle interactions which can include
viscous effects. In this paper we do not need to make explicit
use of these apart from
their production of an impulsive interaction (see section \ref{Uranus} ) 
so  they will  be left  unspecified.  The full
non linear Lagrangian variation is retained for $\psi'$ and ${\bf F}$ as
these may involve the density variation. Contributions coming from the
variation of the central planet potential are included on the left hand side
of Eq.~(\ref{pr}).

\subsubsection{Replacement of convective derivatives}

Lynden-Bell  and Ostriker (1967)
 showed that to first order in the  perturbations,
 the operator
$\Delta$ and the convective operator $(d/dt)$ commute. But note that the
fact that the time derivative is taken at constant ${\mbox{\boldmath$r$}}_0$
taken together with the definition (\ref{Lagrdef}) means that this is true
in general.  Thus for a given quantity $Q$ we have: 
\be 
\Delta \left (dQ
\over dt \right) = {d \over dt} \left( \Delta Q \right) .
\ee 

For quantities $Q$ not involving radial gradients we may linearize and thus
use the unperturbed velocity field ${\bf u}_0({\bf r}_0,t) = (0, r_{0}
\Omega_{r_0} ).$ in working out the convective derivative. We can then write:
\be 
{d \over dt } \left (\Delta Q \right) = { d_0 \over dt } \left ( \Delta
Q \right), 
\ee 
where: 
\be {d_0 \over dt} \left ( \Delta Q \right) = {\partial Q \over \partial t}
+ {\bf u}_0({\bf r}_0,t)\ .\ \nabla Q 
\ee 
Thus, dropping the subscript, we can write the convective derivative ${d
\over dt}$ following the unperturbed motion, and for any quantity $Q$ is: 
\be
{dQ \over dt} = {\partial Q \over \partial t} + \Omega \left( {\partial Q
\over \partial \theta_0} \right) . \label{convd} 
\ee 
Similarly
\be
{d^2Q \over dt^2} = {\partial^2 Q \over \partial t^2} + 2\Omega \left( {\partial^{2} Q
\over \partial t \partial \theta_0 } \right) +
\Omega^2 \left( {\partial^{2} Q
\over \partial \theta_0^2 } \right) . \label{conv2d}
\ee

\subsection{Surface density perturbation and Lagrangian variation 
of the satellite potential} 

We suppose the ring particles to be in eccentric orbits and combine to form
a globally eccentric ring. This is described using a surface density
distribution $\Sigma( r,\theta )$ and eccentricity distribution $e(r).$ We
also consider there to be an axisymmetric reference state for which $e(r) =
\xi_r /r_0 $ and from which we can regard the eccentric ring as being the
result of a perturbation. The perturbation of the surface density is of the
form: 
\be \Sigma( r, \theta)\rightarrow \Sigma( r, \theta) + \Sigma'( r,
\theta). 
\ee 
For linear perturbations $\Sigma' \propto \cos / \sin (m\theta),$ where the
azimuthal mode number, $m = 1.$ The eccentric ring can be thought of as being
primarily in a mode with azimuthal mode number $m=1.$ In practice we may
assume $|e| << 1.$ 
        	
In addition we suppose the ring to be perturbed by a satellite
which produces a contribution to $\psi_s$ of the form: \be \psi_s =
\psi_m(r) \cos(m\theta -m\omega_s t), \label{POT} \ee for general $m$ with
$\omega_s$ being the satellite's orbital frequency.

Note that the analysis presented below may be generalized to
include more than one such term by linear superposition. Expressing the
gradients of $\psi_s$ in cylindrical coordinates, $(r_0, \theta_0)$, gives:
\be 
\Delta \left( {\partial \psi_s \over \partial r} \right) = {\partial
\psi_{s0} \over \partial r_0} + {\mbox{\boldmath$\xi$}}\cdot \nabla \left(
{\partial \psi_{s0} \over \partial r_0}\right) -{\xi_\theta \over r_0^2}
{\partial \psi_{s0} \over \partial \theta_0} \label{fr}
\ee 
and 
\be \Delta \left( {1\over r} {\partial \psi_s \over \partial \theta} \right) = {1\over
r_0}{\partial \psi_{s0} \over \partial r_0} + {\mbox{\boldmath$\xi$}}\cdot
\nabla \left({1\over r_0} {\partial \psi_{s0} \over \partial
\theta_0}\right) +{\xi_{\theta} \over r_0}{\partial \psi_{s0} \over \partial
r_0} \label{fth} 
\ee
\noi where we denote $\psi_{s0} = \psi_s(r=r_0, \theta =\theta_0).$
Note too that we include the components of the gradient of the satellite
potential prior to application of the displacement as the first terms in
(\ref{fr}) and (\ref{fth}). This is because these were not included in the
treatment of the unperturbed axisymmetric ring. 

\subsection{Displacements and couplings in the ring}

Associated with each azimuthal mode number is a displacement
$(\xi_{(m,r)} , \xi_{(m,\theta) }).$ Such a displacement is excited by the
direct action of the satellite acting through the first terms in
Eq.'s~(\ref{fr}) and~(\ref{fth}). This part of the response is that
associated with a satellite in circular orbit acting on a ring in which the
particles are also on circular orbits. In appendix 1
we show why this interaction does not lead to the development of eccentricity
in the ring.

In addition, when it is present, the eccentric mode provides an independent
displacement in the ring with $m=1.$ In a treatment that is fully linearized
in all displacements and exciting potentials this could be superposed with
that excited directly by the satellite.

However, at  an order 
which is the product of the $m=1$ displacement
and the satellite potential
$\psi_{s0},$  effective forcing arising from terms depending on
the product of the displacement component with $m=1$ and gradients of
$\psi_{s0}$ in Eq.'s~(\ref{fr}) and~(\ref{fth}) generates displacement
responses with azimuthal mode numbers $m-1$ and $m+1.$ One or both of these
may produce a resonant response and be important for the dynamics.
This is because such a resonant response can recouple back through the potential
to  excite the original $m=1$ mode resulting in a feedback process,
 
\noindent  Here, we 
focus on the component with $m+1$ and its associated displacement $
(\xi_{(m+1,r)} , \xi_{(m+1,\theta) }).$ However, the component with $m-1$
may be considered by linear superposition if that is important. Thus, with a
single subscript denoting the azimuthal mode number, we consider a general
displacement of the form: \be {\mbox{\boldmath$\xi$}} =
{\mbox{\boldmath$\xi$}} _1 + {\mbox{\boldmath$\xi$}}_{m} +
{\mbox{\boldmath$\xi$}_{m+1}}. \label{PPPP}\ee
\noi The first term represents the displacement associated with $m=1$ mode
in the ring that accounts for the observed eccentricity. The second term is
the displacement directly excited by the circular orbit satellite forcing
potential and the third term is the displacement excited through the
coupling of the satellite potential to the $m=1$ displacement.

\section{The m=1 (eccentric) mode}
\label{m1m}

We here consider the $m=1$ mode which causes the ring to be eccentric. In
the inertial frame the pattern associated with this mode rotates at a low
frequency, $\Omega_p,$ characteristic of the orbital precession frequency
$\omega_{prec} << \Omega $ and the natural time scale of variation is
$\omega_{prec}^{-1}.$ Thus: \be {\partial \over \partial t} \ll \Omega \left(
{\partial \over \partial \theta_0}\right ) \label{ineq}\ee Recalling that
the left hand side of Eq.~(\ref{pth0}) approximated by the linearized
form, gives for the azimuthal component of the displacement:
\be {d \xi_\theta \over dt} + 2\Omega \xi_r   = Q_{\theta_0} \label{qth}\ee

\noi where the quantity $Q_{\theta_0}$ is defined by:
\be {\partial Q_{\theta_0} \over \partial t} + \Omega {\partial Q_{\theta_0}
\over \partial \theta_0}= f_{\theta} - \Delta \left({1\over r} {\partial
\psi' \over \partial \theta } \right). \label{pth}\ee

\noi Using  (\ref{ineq}) gives the adequate approximation:
\be \Omega {\partial Q_{\theta_0} \over \partial \theta_0}= f_{\theta} -
\Delta \left( {1\over r} {\partial \psi' \over \partial \theta } \right).
\label{poth}\ee

Firstly we comment that linear perturbations are strictly separable.
In that approximation,
the $m=1$ component of the
displacement satisfies (Shu\etal 1985):
\be {\partial^2 \xi_r\over \partial \theta_0^2 } = - \xi_r. \ee

Secondly we comment that the motion is dominated by the force due
to  the central mass, and to 
within an error governed by  the magnitude of the other forces acting
in comparison, is
Keplerian. This means that to within this limit
which is also measured by the ratio of orbital precession 
frequency to orbital frequency
\be { \partial
\xi_\theta \over \partial \theta_0 } = - 2 \xi_r \label{Kep}\ee which
applies to Keplerian orbits with small eccentricity.

\noindent Furthermore ({\ref{qth}) tells us
 that the magnitude of $Q_{\theta_0}$ compared to that of
$\Omega \xi_r$  is of order the ratio of 
 forces producing non Kepler motion to the force due to the central mass. 

Using (\ref{qth}) and the assumption of slow rate of change to neglect
second partial derivatives with respect to time in Eq.~(\ref{pr}) one
then finds that the $m=1$ component of the displacement satisfies:
\be 2\Omega {\partial^2 \xi_r\over \partial t \partial \theta_0} - \xi_r
(\Omega^2 - \kappa^2 )= f_{r} - \Delta \left( {\partial \psi' \over
\partial r} \right) + 2\Omega Q_{\theta_0},\label{rmot}\ee

\noi Here the square of the epicyclic frequency is given by: \be \kappa^2
={2\Omega \over r_0}{d (r_0^2 \Omega)\over dr_0}. \label{epi}\ee

To evaluate the Lagrangian change to the gradient of the satellite potential
we insert Eq.~(\ref{PPPP}) in Eq.~(\ref{fr}). However, we retain
only the terms with azimuthal dependence corresponding to $m=1.$ The
assumption here is that only these produce significant secular effects on
the mode. Other terms produce small corrections with a different and in
general much more rapidly varying azimuthal dependence.
       
Thus we adopt  the following combination of terms from (\ref{fr}):
\be \Delta_{sec} \left( {\partial \psi_s \over \partial r} \right) =
{\mbox{\boldmath$\xi_{m+1}$}}\cdot \nabla \left( {\partial \psi_{s0} \over
\partial r_0}\right) -{\xi_{m+1,\theta} \over r_0^2} {\partial \psi_{s0} 
\over \partial \theta_0}, \label{fr1}\ee 
where $\Delta_{sec}$ refers to terms that can contribute secularly to the $m=1$
mode.  Note that because ${\partial \psi_s / \partial r_0}$ has azimuthal
mode number $m,$ it is not included. Notice that the retained terms are the
only non-vanishing ones when multiplied by $\cos / \sin {(\theta_0)}$ and
integrated over $\theta_0.$

\section{Conservation of radial action} 
\label{RA}

As is well known, the radial action, taken here to be $\sqrt{GM_*a}
(1-\sqrt{1-e^2}),$ per unit mass associated with a Keplerian orbit around
the central mass, $a$ here being the semi-major axis is an adiabatic
invariant.  As the ring particles  are always close to eccentric Keplerian
orbits, we might expect to find a related quantity. 

\noi To do this we define:
\be I_r = \int \Sigma_0 \Omega \left( {\partial \xi_r\over \partial
\theta_0}\right)^2 r_0 dr_0 d\theta_0 \label{edeff}.\ee Here and in other similar
integrals are taken over the entire radial and azimuthal domain of the ring.

To see that for $m=1$ displacements the above integral corresponds
to the radial action we note that if we identify $a \equiv r_0$ and use the
fact that: \be {1\over 2\pi} \int_0^{2\pi} (\xi_r/r_0)^2 d\theta_0 \equiv
{1\over 2} e^2, \label{edeff1} \ee $I_r$ may be written:
\be I_r = {1\over 2} \int a^{1/2} e^2 d\mu ,\ee with $d\mu$ being a mass
element associated with the ring.  This corresponds to a multiple
$1/\sqrt{GM_*}$  of the
standard radial action expanded to first order in $e^2.$
 
\subsection{ Rate of change of radial action for the ring}\label{RACT}
We first comment that self-gravity makes no contribution
(see appendix 2 for details) and so when calculating the rate
of change of radial action in this section  we may replace $\psi'$ by $\psi_s.$

\noindent By multiplying (\ref{rmot}) by ${\partial \xi_r / \partial \theta_0}$ and
integrating over the mass of the ring we obtain an expression for the time
rate of change of the radial action in the form:
\be 
{\partial \over \partial t} \left( \int \Sigma_0 \Omega \left({\partial
\xi_r\over \partial \theta_0}\right)^2 r_0 dr_0 d\theta_0\right) =\int
\Sigma_0 {\partial \xi_r\over \partial \theta_0} \left(f_{r} -
{\mbox{\boldmath$\xi_{m+1}$}} \cdot \nabla \left( {\partial \psi_{s0} \over
\partial r_0}\right) +{\xi_{m+1,\theta} \over r_0} {\partial \psi_{s0} 
\over \partial \theta_0} +2\Omega Q_{\theta_0}\right) r_0 dr_0 d\theta_0
\label{consv}
\ee
\noi We do an integration by parts with respect to $\theta_0$ making
use of (\ref{poth}) together with (\ref{fth}) This procedure enables us to
eliminate $Q_{\theta_0}$ in terms of the satellite potential. Then
(\ref{consv}) becomes:
$$ {\partial \over \partial t} \left(
\int \Sigma_0 \Omega \left({\partial \xi_r\over \partial \theta_0}\right)^2
r_0 dr_0 d\theta_0\right) =\int \Sigma_0 {\partial \xi_r\over \partial
\theta_0} \left(f_{r} - {\mbox{\boldmath$\xi_{m+1}$}} \cdot \nabla \left(
{\partial \psi_{s0} \over \partial r_0}\right) +{\xi_{m+1,\theta} \over
r_0^2} {\partial \psi_{s0} \over \partial \theta_0} \right) r_0 dr_0
d\theta_0 $$ 
\be 
+\int \Sigma_0 {\partial \xi_{\theta} \over \partial
\theta_0}\left( f_{\theta} - {\mbox{\boldmath$\xi_{m+1}$}}\cdot \nabla
\left({1\over r_0} {\partial \psi_{s0} \over \partial \theta_0} \right)
-{\xi_{m+1,\theta}\over r_0} {\partial \psi_{s0} \over \partial r_0}
\right)\ r_0\ dr_0\ d\theta_0\ \label{consv1}
\ee

\noindent Eq.~(\ref{consv1}) gives a conservation law for the radial action
associated with the $m=1$ mode.

In the absence of collisional terms involving $f_r, f_\theta $ and terms
involving the satellite potential, as expected the radial action is
conserved. It is important to note that the collisional terms act on the {
\it perturbation from an axisymmetric state} and are not necessarily
negative definite leading to a decay of the radial action ( see eg.
Papaloizou  and Lin 1988 ) though in general one might expect that to be the
case. The terms on the right hand side of (\ref{consv1}) that depend on the
m-component of the satellite potential can also change the radial action
when the satellite forcing leads to secular changes due to waves launched at
resonance (see eg. Goldreich  and Tremaine 1980).

For a ring in which a steady eccentricity is maintained the secular rates of
change due to satellite forcing must balance those due to the effects of viscosity or 
collisions acting on the $m=1$ perturbations. In such a case we have a
condition such that $I_r$ remains constant. 

Notice that there is no net-contribution from self-gravity when integrated
over the mass of the ring, hence, any dissipation arising from internal
friction can only be compensated by external forces, i.e. the satellite
torques (see also the discussion below). 
The exact fraction of the satellite torque that acts on the ring so
as to maintain the $m=1$ mode is calculated in the next section.  

\section{The satellite Torque}
\label{SATT}

The satellite potential terms arising on the right-hand side of
Eq.~(\ref{consv1}) are directly related to the resonant torque
generated by the $m+1$ forcing between satellite and ring as occurs in
Eq.'s~(\ref{pr}) and~(\ref{pth0}) through the forcing gradient of
potential components: 
\be 
\left( \Delta \left( {\partial \psi_s \over \partial r} \right)
\right)_{m+1} = {\mbox{\boldmath$\xi$}}\cdot \nabla \left( {\partial
\psi{s_0} \over \partial r_0} \right) -{\xi_\theta \over r_0^2} {\partial
\psi_{s_0} \over \partial \theta_0} \label{frs}\ee and \be \left( \Delta
\left( {1\over r} {\partial \psi_s \over \partial \theta} \right)
\right)_{m+1} = {\mbox{\boldmath$\xi$}}\cdot \nabla \left({1\over r_0}
{\partial \psi_{s_0} \over \partial \theta_0} \right) +{\xi_{\theta} \over
r_0}{\partial \psi_{s_0} \over \partial r_0}, \label{fths} 
\ee 
\noi where the terms not containing the $m+1$ components are ignored
thus the unsubscripted displacement is the $m=1$ component. Note that the
forcing amplitude is proportional to both the satellite potential and the
ring eccentricity. Note that in principle additional forcing terms can arise
from the components of $( f_r, f_{\theta})$ with azimuthal mode number $m +
1.$ Such components can for example be generated from terms proportional to
the ring eccentricity and those coming from the response to the satellite
forcing potential component with azimuthal mode number $m.$ However, this
response is non-resonant in the neighborhood of the $m+1$ Lindblad
resonance where the response to $m+1$ forcing is located. Forcing terms of
this type are not considered here as they arise from forces internally
generated in the ring, which are assumed small compared to those arising
from direct forcing. Furthermore we only require the ring-satellite torque
produced by the direct forcing of the unperturbed background ring through
Eq.s~(\ref{frs}) and~(\ref{fths}) which from the above discussion
should represent almost the total torque.

The total torque  can be estimated as in Goldreich  and Tremaine (1978). To
provide an expression for this torque, we return to the equations of
motion~(\ref{pr}) and~(\ref{pth0}). From these the rate of change of ring
canonical energy (see Freidman  and Schutz 1978) associated with the response
with azimuthal mode number $m+1$ is obtained as the rate of doing work as a
result of the forcing as:
\be 
{\dot E}_{m+1} = -\int\ \Sigma\ \left({\partial \xi_{m+1, r} \over
\partial t }\ \left( \Delta\left({\partial \psi_s \over \partial r} 
\right)\right)_{m+1} + {\partial \xi_{m+1, \theta} \over \partial t}\
\left(\Delta \left( {1 \over r} {\partial \psi_s \over \partial \theta }
\right) \right)_{m+1} \right)\ r\ dr\ d\theta. \label{Edot} 
\ee

We now use the fact that because the disk is forced by a disturbance with a
definite pattern speed, $\Omega_{PP},$ say the rate of change of ring
angular momentum is given by: \be {\dot J}_{m+1} = { {\dot E}_{m+1} \over
\Omega_{PP}} , \ee which expresses the well known result that the ratio of 
energy to angular momentum exchanged is $\Omega_{PP}$ (see also Freidman and 
Schutz 1978).
        
\noi Similarly for the perturbation itself we have: 
\be {\partial
\xi_{m+1, r} \over \partial t } = - \Omega_{PP} {\partial \xi_{m+1, r} 
\over \partial \theta }. \label{BBBB} 
\ee 

Taking Eq.'s~(\ref{Edot}) and~(\ref{BBBB}) together we obtain for the
rate of change of ring angular momentum: 
\be 
{\dot J}_{m+1} =\int\ \Sigma\ \left({\partial \xi_{m+1, r} \over
\partial \theta }\ \left( \Delta\left({\partial \psi_s \over \partial r}
\right)\right)_{m+1} + {\partial \xi_{m+1, \theta} \over \partial \theta}\
\left(\Delta \left( {1\over r} {\partial \psi_s \over \partial \theta }
\right) \right)_{m+1} \right)\ r\ dr\ d\theta. 
\ee
\noi Because the above are second order expressions in the
perturbations, we may introduce the subscript $0$ into $r$ and $\theta.$

Using (\ref{frs}) and (\ref{fths}) and then noting that as the
expressions are second order in the perturbations, we can  apply the
subscript $0$ to $r,$ $\theta ,$ and the background surface density so as
to obtain:
$$ {\dot J}_{m+1} =-\int  
\Sigma_0\ {\partial \xi_{m+1, r} \over \partial \theta } \left(
-{\mbox{\boldmath$\xi$}}\cdot \nabla \left( {\partial \psi_{s0} \over
\partial r_0} \right) +{\xi_\theta \over r_0^2} {\partial \psi_{s0} \over
\partial \theta_0} \right)\ r_0\ dr_0\ d\theta_0$$ 
\be 
-\int \Sigma_0\
{\partial \xi_{m+1, \theta} \over \partial \theta}\left(
-{\mbox{\boldmath$\xi$}}\cdot \nabla \left({1\over r_0} {\partial \psi_{s0}
\over \partial \theta_0} \right) -{\xi_{\theta}\over r_0}{\partial \psi_{s0}
\over \partial r_0} \right)\ r_0\ dr_0\ d\theta_0 . \label{TOR} 
\ee

\section{The Radial action and the satellite Torque}
\label{RAST}

We write the satellite potential as in Eq.~(\ref{POT}) and the
displacements as:
\begin{eqnarray*}
\xi_r  &=& A_r\  \cos(\theta_0 - \Omega_P\ t)\\ 
\xi_{\theta} &=&  A_{\theta}\ \sin(\theta_0 - \Omega_P\ t) \\
\xi_{m+1,r} &=& A_{m+1,r}\ \cos((m+1)\theta_0 - (\Omega_P + m\ \omega_s)\ t + \beta(r_0))\\ 
\xi_{m+1,\theta} &=& A_{m+1,\theta}\ \sin((m+1)\theta_0 - 
      (\Omega_P + m\ \omega_s)\ t + \beta(r_0)) 
\end{eqnarray*}
Where $\Omega_P$ is the slow pattern frequency or precession frequency
associated with the $m=1$ mode as seen in the inertial frame. $\beta(r_0)$ is
the azimuthal phase shift undergone by the wave crests between the
resonance location and $r_0$ (see for example Goldreich  and Tremaine 1978).
Notice that for the $m=1$ mode this is a very small quantity that has been
neglected. The amplitude of the $m+1$ mode, $A_{m+1}$, is a function of the
satellite forcing.

We also assume that the disturbance
${\mbox{\boldmath$\xi_{m+1}$}}$ is generated and significant only near a
Lindblad resonance. For the case with azimuthal mode number $m+1$ this is
where: 
$$(m+1)\Omega - \Omega_P - m\ \omega_s = \pm \kappa .$$ 
We note that the pattern speed is $\Omega_{PP} =(\Omega_P +
m\omega_s)/(m+1)$ and the necessary choice of negative sign, to enable the
resonance to lie within the ring means that it corresponds to an outer
Lindblad resonance. This occurs where the ring rotates more slowly that the
satellite and, accordingly, gains angular momentum.

Were we dealing with the case of azimuthal mode number $m-1,$ the
choice of positive sign would be necessary and we would have an inner
Lindblad resonance at which the ring lost angular momentum.

To within an error of order
the ratio of the precession to orbital
frequency (typically $10^{-4}$ for the 
$\epsilon$ ring around Uranus; see section {\ref{discu})
we may replace $\kappa$ by $\Omega$ and use
the relation between $\xi_{m+1,r}$ and $\xi_{m+1,\theta}$ that applies at
and in the neighborhood of a Lindblad resonance.  To determine this
relation, internal forces arising in the ring and from any satellite may be
neglected. Eq.~(\ref{pth0}) then reduces to: 
\be {d \xi_\theta \over
dt}+ 2\Omega \xi_r = 0. \ee Using Eq.~(\ref{convd}) together with
Eq.~(\ref{BBBB}) we find: \be ( \Omega - \Omega_{PP}){\partial 
\xi_{m+1,\theta} \over \partial \theta_0} +2\Omega \xi_{m+1,r} =0.
\ee
Because we are close to resonance, we use the above condition for an outer
Lindblad resonance to get $ (m+1)( \Omega - \Omega_{PP}) = -\Omega$ so
obtaining: 
\be {\partial \xi_{m+1,\theta} \over \partial \theta_0} -2(m+1) \xi_{m+1,r}
=0. \label {OLRC}
\ee
\noi This is very similar to Eq.~(\ref{Kep}) which reads
\be 
{ \partial \xi_\theta \over \partial \theta_0 } = - 2 \xi_r \label{Kop}
\ee 
From (\ref{OLRC}) and (\ref{Kop}) we find the useful result that 
\be
{A_{m+1,\theta}\over A_{m+1,r}} = - {A_{\theta}\over A_{r}} 
\ee 
Then, the non vanishing terms in Eq.~(\ref{TOR}) lead to: 
\be {\dot J}_{m+1}
=-\int \Sigma_0\ C_T(r_0) \sin({\beta(r_0)})\ r_0\ dr_0\ d\theta_0.
\label{SAT_TOR} 
\ee 
The above coefficients are given by: 
\be 
C_{T}(r_0)\ =  \frac{(m+1) A_{m+1,r}A_{r}}{4 } \
{\partial^2 \psi_{s0} \over \partial r_0^2} -\frac{ (m+1)
A_{m+1,\theta}A_{\theta}}{4 r_0 } \left( m^2\ \frac{\psi_{s0}}{r_0}\ -\
{\partial \psi_{s0} \over \partial r_0}\right).\\ 
\ee 
It turns out that the terms associated with the satellite-potential which
ultimately give a non vanishing contribution in Eq.~(\ref{consv1})
which gives the rate of change of ring radial action can be expressed as
multiple of Eq.~(\ref{SAT_TOR}). Thus:
\begin{eqnarray*}
\left[ {\partial \xi_r\over \partial \theta_0}
\left( {\mbox{\boldmath$\xi_{m+1}$}}\cdot \nabla \left( {\partial
\psi_{s0} \over \partial r_0} \right) 
- {\xi_{m+1,\theta} \over r_0^2} {\partial \psi_{s0} \over \partial
\theta_0} \right)\ \right]_{sec}\ + &  & \\ 
+\ \left[ {\partial \xi_{\theta} \over \partial \theta_0}\ \left( 
{\mbox{\boldmath$\xi_{m+1}$}}\cdot \nabla \left({1\over r_0} {\partial
\psi_{s0} \over \partial \theta_0} \right) + 
{\xi_{m+1,\theta}\over r_0} {\partial \psi_{s0} \over \partial r_0}
\right)  \right]_{sec}\ & = &  C_{RA}(r_0)\ .\ \sin({\beta(r_0)}), 
\label{SAT_RA}
\end{eqnarray*}
\noi where the subscript $sec$ here denotes an azimuthal average and it is
verified that:
$$ C_{RA}(r_0) = C_{T}(r_0)/(m+1).$$
\noi Thus, Eq.~(\ref{consv1}) can be re-written as:
\begin{eqnarray*}\label{consv2}
{\partial \over \partial t} \left( \int\ \Sigma_0\ \Omega \left({\partial
\xi_r\over \partial \theta_0}\right)^2\ r_0\ dr_0\ d\theta_0 \right) &=&
\int\ \Sigma_0\ \left( {\partial \xi_r\over \partial \theta_0}\ f_{r} +
{\partial \xi_{\theta} \over \partial \theta_0}\ f_{\theta} \right)\ r_0\
dr_0\ d\theta_0\\ &+&\frac{{\dot J}_{m+1}}{m+1}. 
\end{eqnarray*}
\noi We can also add in the effects of forcing with azimuthal mode number $m-1$
should that lead to secular effects by replacing ${\dot J}_{m+1}/(m+1)$ by
$$ \frac{{\dot J}_{m+1}}{m+1} + \frac{|{\dot J}_{m-1}|}{m-1} .$$

Note that the absolute value of ${\dot J}_{m-1}$ appears as the case of
azimuthal mode number $m-1$ corresponds to an inner Lindblad resonance for
which ${\dot J}_{m-1} < 0.$
       
We may then write Eq.~(\ref{consv1}) in the compact form applicable to
a thin ring: 
\be {d I_r \over dt} = -{{\dot E_d}/\Omega} + \left( \frac{{\dot
J}_{m+1}}{m+1} + \frac{|{\dot J}_{m-1}|}{m-1} \right ) . \label{compact}
\ee 
Here $\Omega$ may be evaluated at the ring center and: 
\be {\dot E_d} = -\int\ \Sigma_0\ \Omega \left( {\partial \xi_r\over \partial
\theta_0}\ f_{r} + {\partial \xi_{\theta} \over \partial \theta_0}\
f_{\theta} \right)\ r_0\ dr_0\ d\theta_0 \label{actf}\\ 
\ee 
is a quantity having the dimensions of the rate of energy dissipation of the
perturbed motion. Eq.~(\ref{compact}) gives a condition for non zero
radial action to remain constant, that requires external satellite torque
input to balance dissipative effects due to particle collisions.

\noindent We comment here in the context of an application below (see section \ref{discu})
that the dissipative term takes a particularly simple form
when the forces per unit mass  $( f_r, f_{\theta})$ are taken
to produce small impulsive changes $(\Delta_I(d\xi_r/dt) , \Delta_I(d\xi_{\theta}/dt))$
in the perturbed motion that may for example take
place once per orbit at pericenter (see Eqs. \ref{pr} and \ref{pth0}). 
Then 
\be  {\dot E_d}= - {1\over 2} \int\Omega \left(\Delta_I(d\xi_r/dt)^2+ \Delta_I(d\xi_{\theta}/dt)^2\right)
  \Sigma_0 \ r_0\ dr_0\ ,\ee
which is the difference in kinetic energy before and after the impulse
as experienced by a fluid element as 
evaluated using the Lagrangian velocity perturbations per orbital period.

However, as is well known there is another condition that has to be
satisfied in order that the ring may possess a steady $m=1$ mode. This comes
about because the ring has to precess at a uniform rate. This requires
internal forces due to self-gravity and particle collisions to balance the
differential precession that would occur if ring particles moved freely
under the central potential (see for example Chiang and Goldreich 200). We
now give an expression for this condition in a similar format to that of
Eq.~(\ref{compact}).

\section{The condition for uniform precession}
\label{UP}

The $m=1$ mode responsible for the ring eccentricity has a constant
and very small pattern speed as viewed in the inertial frame. This
means that individual ring particles appear to be in elliptic orbits
that precess at the same rate. In oder to achieve this the internal
and external forces acting in the mode have to satisfy a constraint
that can be view as a non-linear dispersion relation. Our treatment
again follows that of Shu\etal (1985) who provided such a relationship
for density waves in Saturn's rings. Except here we consider a density
wave comprising a global normal mode rather than a forced propagating
wave.

Eq.~(\ref{pr}) can be  expressed in the form:
\be 
{d^2 \xi_r\over dt^2} + \xi_r \kappa^2 = f_{r} - \Delta \left({\partial
\psi' \over \partial r} \right) + 2 \Omega Q_{\theta_0} \label{pr2}
\ee 
We now use an angle that is fixed with respect to a coordinate system
rotating at the pattern angular frequency $\Omega_P,$ namely $\phi_0 =
\theta_0 - \Omega_P\ t.$ The radial displacement is taken to be of the form
$\xi_r = A(r_0)\ \cos(\phi_0).$ Following Shu\etal (1985) we note that as the
time dependence is contained within $\phi_0,$ $\xi_r$ only depends on $r_0$
and $\phi_0.$
       
Multiplying Eq.~(\ref{pr2}) by $\cos(\phi_0)$ and integrating over
$\phi_0,$ we obtain: 
\be 
{1\over 2}\left( \frac{\kappa^2}{(\Omega - \Omega_P)^2} - 1
\right)\ A(r_0) = \frac{1}{(\Omega - \Omega_P)^2}\left( F_{cr} + g_D(r_0)\\
+ \frac{1}{2\pi} \int_0^{2\pi} 2\ \Omega Q_{\theta0} \cos(\phi_0) d\phi_0
\right), \label{pr3} \ee where: \be F_{cr} = \frac{1}{2\pi} \int_0^{2\pi}
f_r \cos(\phi_0) d\phi_0, \ee and \be g_{D}(r_0) = - \frac{1}{2\pi}
\int_0^{2\pi} \cos(\phi_0) \Delta \left({\partial \psi_{SG} \over \partial
r} \right) d\phi_0 
\ee 
We note here that the satellite potential does not contribute to the
determination of the $m=1$ mode to  within an error
on the order of the ratio of the satellite forcing potential
to that due to the central mass and so is neglected in this
section.

The last term in Eq.~(\ref{pr3}) can be re-written after an integration
by parts as in section~\ref{RA} in the form: 
\be 
\frac{1}{2\pi} \int_0^{2\pi} 2\ \Omega Q_\theta \cos(\phi_0) d\phi_0 = 2\
F_{c\theta} , \ee where: \be F_{c\theta} = - \frac{1}{2\pi} \int_0^{2\pi}
f_\theta \sin(\phi_0) d\phi_0. 
\ee
Here we have  assumed  that the azimuthal scale is much longer than the radial one
so that the azimuthal component of the acceleration due
to self-gravity can be neglected in comparison to the radial one
(tight winding approximation).

\noindent Eq.~(\ref{pr3}) then becomes: 
\be {1\over 2} \left( \frac{\kappa^2}{(\Omega - \Omega_P)^2} - 1 \right)\ A(r_0) =
\frac{g_{ext}}{(\Omega - \Omega_P)^2}\, \label{pr4} \ee where: \be g_{ext} =
\left( F_{cr} + 2\ F_{c\theta} \right) + g_D . \label{gext} 
\ee 
Given that $\kappa = \Omega - \omega_{prec}$, where $\omega_{prec}(r_0)$ is
the local radius dependent precession frequency and assuming that $\Omega_P
<< \Omega$ and $\omega_{prec} << \Omega$, Eq.~(\ref{pr4}) can be
approximated to first order in $\Omega_P$ and $\omega_{prec}$ as: 
\be 
(\Omega_P - \omega_{prec})\ A(r_0) = \frac{g_{ext}}{\Omega}
\label{pr5} 
\ee 
Eq.~(\ref{pr5}) provides a condition to be satisfied by the normal mode
amplitude that can be thought of as a condition for uniform precession. It
is satisfied when the self gravity, satellite forcing and the internal
collisional terms represented on the right-hand side of Eq.~(\ref{pr5})
balance the differential precession term on the left-hand side. 
Note further that for a thin ring of the type considered
here, $\Omega$ may be taken as constant in (\ref{pr5})
and evaluated
at the ring center from now on.

\section{The self-gravity term}
\label{SGterm}

In order to calculate $g_D$ we follow Shu (1985). As radial variations are
much more rapid than azimuthal ones, the local self-gravity at $r_0$ is
canonically approximated to be that due to an infinite plane sheet of radial
width $\Delta r = r_2 - r_1$, where $r_1$ and $r_2$ are the inner and outer
bounding radii of the  ring beyond which 
the surface density vanishes  respectively. Thus: 
\be
\left({\partial \psi_{SG} \over \partial r} \right) = {2G\over {\overline r}} \ \int_{r_1}^{r_2}
\frac{\Sigma(r')}{ (r-r')} r'dr', \label{SGF}
\ee
where ${\overline r}$ is the mean radius of the ring
 and $G$ is the gravitational constant.

\noindent Note that we use $\Sigma(r') r'/{\overline r}$ 
in the above rather than
$\Sigma(r').$ Although the difference is apparently not significant
in the slender ring approximation, we use the prescription
we do to ensure that gravity remains conservative under the tight winding
approximation with gravitational energy
\be {\cal U} = {2G\over {\overline r}}  \int_{r_{1}}^{r_{2}}
\ln|r_0-r_0'|
\Sigma(r_0') \Sigma(r_0)r_0'r_0 dr_0'dr_0 .\ee

\noindent  Because of the small mass of the ring, self-gravity
is neglected in the axisymmetric background state,
consistently with that Eq. (\ref{SGF}) gives the  Lagrangian variation  also.
Thus
\be
\Delta
\left({\partial \psi_{SG} \over \partial r} \right) = {2G\over {\overline r}} \ \int_{r_1}^{r_2}
\frac{\Sigma(r')}{ (r-r')} r'dr'
.\label{gs1}
\ee
Note that were the background axisymmetric contribution to
self-gravity incorporated here it would make no difference to
the subsequent analysis as it azimuthally averages to zero.

In addition, $r = r_0 + \xi_r$ and $r' = r_0' +
\xi_r'$, where $\xi_r = \xi_r(r_0) = A(r_0)\ cos(\phi_0) $ and $\xi_r' =
\xi_r(r_0') = A(r_0')\ cos(\phi_0).$ In this planar limit, we identify the
ring eccentricity as $e(r_0) = A(r_0)/{\overline r} .$ Using the tight-winding
approximation we have: 
\be 
\Sigma(r')r'dr' = \Sigma(r_0')r_0'dr_0' 
\ee 
which represents conservation of mass. We then  have: 
\be 
\Delta \left({\partial \psi_{SG} \over \partial r} \right) = {2G\over {\overline r}}
\int_{r_1}^{r_2} \frac{\Sigma(r_0')}{ r_0 + \xi_r - r_0' - \xi_r'}r_0' dr_0'
\label{gs2} 
\ee 
We can re-write Eq.~(\ref{gs2}) in terms of the eccentricity gradient,
$q$:
\be q= \frac{A(r_0) - A(r_0')}{r_0 - r_0'}. 
\label{q} 
\ee 

\noindent Then after integrating over $\phi,$ we obtain (see also Shu\etal 1985): 
\be g_{D} = 2G\ \int_{r_1}^{r_2} \frac{I(q)}{q} \Sigma(r_0')\ \frac{A(r_0) -
A(r_0')}{(r_0 - r_0')^2} dr_0' \label{gs21} 
\ee 
where: 
\be 
I(q) = \frac{1}{2\pi} \int_0^{2\pi} \frac{cos(\phi)}{ 1 - q\cos{\phi}}
d\phi = \frac{1}{q \sqrt{1 - q^2}}\ \left(1 - \sqrt{1 - q^2}\right) 
\ee
and, to within a  small error of order $\Delta r/{\overline r},$
 we have replaced $r_0'dr_0'$ by ${\overline r}dr_0'.$
\noindent Notice that the integrand in Eq.~(\ref{gs21}) presents a singularity,
that has to be handled in a principal value sense,
which can lead to practical complications near the ring edges.  

\noindent With that comment in mind we note that  when Eq.(\ref{pr5})
 is combined with Eq.(\ref{gs21}) to give a single
equation, viewed as condition for uniform
precession of the ring from which one can determine $A(r_0),$
one has a continuum form  of Eq.(14) of Goldreich and Tremaine (1979)
which gives a condition for uniform precession
in terms of $N$ discrete constraints on $N$ elliptical streamlines.

\section{The value of q}
\label{vq}

If we assume that $q$ is constant throughout the ring, then, from
Eq.~(\ref{gs21}), it can be shown that: 
\be \label{q1} \int_{r_1}^{r_2}\
g_D\ \Sigma_0(r_0)\ A(r_0) dr_0 = G\ \frac{I(q)}{q}\int_{r_1}^{r_2}
\int_{r_1}^{r_2} \Sigma_0(r_0)\ \Sigma_0(r_0')\ \left(
\frac{A(r_0)-A(r_0')}{r_0 - r_0'} \right)^2 dr_0 dr_0'\ > 0 \ee

On the other hand, from Eq.~(\ref{pr5}) we can write:
\be
\label{q2}
\int_{r_1}^{r_2}\ (\Omega_P - \omega_{prec})\ A(r_0)^2\ \Sigma_0(r_0)\ dr_0
= {1\over \Omega} \int_{r_1}^{r_2}\ g_{ext}\ \Sigma_0(r_0)\ A(r_0) dr_0 
\ee

Thus, if the collisional impulses can be neglected with respect to the
self-gravity, it is verified on setting $g_{ext} = g_D$ in Eq. (\ref{q2})  that:
\be
\label{q3}
\int_{r_1}^{r_2}\ (\Omega_P - \omega_{prec})\ A(r_0)^2\ \Sigma_0(r_0)\ dr_0
> 0
\ee

We remark that the pattern speed, $\Omega_P$ may be regarded as an
eigenvalue associated with a normal mode determined by Eq.~(\ref{pr5}).
By multiplying Eq.~(\ref{pr5}) by $ \Sigma(r_0)$ and integrating
over the ring it follows that $\int^{r_1}_{r_2} (\Omega_P - \omega_{prec})A(r_0)\Sigma_0r_0dr_0 = 0.$
Hence  if $A(r_0)$ does not change sign, as in the
case of interest here,  $\Omega_P$ must equal the local precession
frequency, $\omega_{prec}$ at some intermediate point in the ring,
$\bar{r_0}$, i.e. $\Omega_P = \omega_{prec}(\bar{r_0})$. 

Then, we can write:
\be
\label{q4}
\int_{r_1}^{r_2}\ (\Omega_P - \omega_{prec})\ A(r_0)\ (A(r_0)-A(\bar{r_0}))\ \Sigma_0(r_0)\ dr_0
> 0,
\ee
which can be put as:
\be
\label{q5}
\int_{r_1}^{r_2}\ (\Omega_P - \omega_{prec})\  A(r_0)^2\
\frac{q}{A(r_0)}\ (r_0 - \bar{r_0})\ \Sigma_0(r_0)\ dr_0
> 0,
\ee
from which we obtain: $-(d \omega_{prec}/dr_0)\ \times q/A(r_0) > 0$. Thus,
$ (1/e)(de/dr_0) $ and q (since $A(r_0)$ does not change sign),
is necessarily positive in any ring where self-gravity is the main mechanism
that maintains apse alignment (see also Goldreich  and Tremaine 1979,
Borderies\etal 1983).

To estimate the value of $q$ in a narrow-eccentric ring, 
in the linear regime where $q<<1$, $2I(q)/q \approx 1$ and if
we neglect all perturbations other than self-gravity, from Eq.'s~(\ref{pr5})
and~(\ref{gs21}), we can write: 
\be
|\Delta \omega_{prec}| = \frac{GM_r|\Delta e |}{2\pi r\Omega e (\Delta r)^2}
\label{q6} 
\ee 
Here $|\Delta \omega_{prec}|$ gives the magnitude of the difference between
the free particle precession frequencies at the ring edges. Hence: 
\be |q| \sim r{ |\Delta e |\over |\Delta r|} = {2\pi e \Omega r^2 \Delta r\Delta
\omega_{prec} \over GM_r} 
\ee 
Hence, the magnitude of $q$ is basically determined by the mass, size and
eccentricity of the ring.

%
%%%%%%%%%%%%%%%%%%%%%%%%%%%%%%%%%%%%%%%%%%%%%%%%%%%%%%%%%%%%%%%%%%%%
\section{Application}
\label{discu}
       
In this paper we have presented a description of a narrow
self-gravitating ring in orbit about a dominant central mass. We have
considered the situation when the ring displays an eccentricity through
sustaining a global non-axisymmetric $m=1$ mode of oscillation.  Such a mode
with a single pattern speed corresponds to an eccentric ring undergoing
uniform solid body precession.
	
We have used a continuum description of the system following that of
Shu\etal (1985) for density waves in Saturn's rings rather than a
description in terms of an ensemble of interacting streamlines. One can in
principle include the effects of self-gravity, viscosity or  particle collisions and
external satellites. In the latter case we do not assume a continuous
distribution of resonant interactions (eg. Borderies\etal 1983) but allow
for a small number of separated resonances.

Two main conditions for the ring to be able to maintain a $m=1$ mode with
single slow pattern speed are obtained. One can be expressed as an integral
condition for the pattern or normal mode to precess at a uniform rate
(Eq.~(\ref{pr4})) that requires the correct balance between 
differential precession, self-gravity and collisional effects. The other
condition is one for the non zero net radial action in the ring, that exists
because of its finite eccentricity, to be sustained through a balance of
injection due to eccentric resonances arising from external satellites and
additional collisional damping associated with the presence of the $m=1$
mode (Eq.~(\ref{actf})).

\subsection{The radial action, the satellite torque and the dissipation: 
A more simple N-body approach}\label{RACTS}

The derivation we presented above based on Fourier decomposition in the
azimuthal direction was rather lengthy. We here look at how the radial
action changes directly in a simpler manner and connect the result to the
behavior of a global $m=1$ mode given by Eq.~(\ref{actf}).

If we consider the ring as consisting of $N$ particles of mass $m_i,$ having
semi-major axes $a_i$ and eccentricities $e_i,$ $i =1,2,3,....$ The total
radial action is: 
\be 
I_r = \sum_{i=1}^{N} m_i\sqrt{GM_*a_i} 
\left(1-\sqrt{1-e_i^2} \right).
\ee 
In terms of the energies $E_i = -GM_*m_i/(2a_i),$ and angular momenta  
$J_i = m_i\sqrt{GM_* a_i (1-e^2_i)},$
 this can be
written: \be I_r = \sum_{i=1}^{N}\left( 2^{-1/2} GM_*(m_i)^{3/2}
\left(-E_i\right)^{-1/2} - J_i \right ). 
\ee 
Accordingly we have:
\be {d I_r \over dt} = \sum_{i=1}^{N}\left( {1\over \Omega_i }{dE_i\over dt}
-{d J_i\over dt} \right ), \label{Irt}
\ee 
with $\Omega_i =\sqrt{GM_*}/(a_i)^{3/2}.$ We now find it convenient to
introduce a fixed semi-major axis $a_0$ which could correspond to the ring
center and rewrite Eq.~(\ref{Irt}) in the form: 
\be {d I_r \over dt} =
\sum_{i=1}^{N}\left( \left({1\over \Omega_i } - {1\over \Omega_0 }\right)
{d( E_i - E_{i0})\over dt} +{1\over \Omega_0 }{dE_i\over dt} -{d J_i\over
dt} \right ), \label{Irt2}
\ee 
where $E_{i0} = -GM_*m_i/(2a_0)$ and $\Omega_0 =\sqrt{GM_*}/(a_0)^{3/2}.$

Now we may use the conservation of total energy and angular momentum to
write:
\be 
\sum_{i=1}^{N} {d J_i\over dt} = {\dot J_s}
\label{Jrt}
\ee
and
\be 
\sum_{i=1}^{N} {d E_i\over dt} + {d {\cal U} \over dt}  = {\dot E_s} - {\dot E_{dissip}},
\label{Frt} 
\ee
where ${\dot J_s}$ and ${\dot E_s}$ are the rate of angular momentum and
energy input into the ring from external satellites and ${\dot E_{dissip}}$
is the rate of dissipation due to inelastic collisions. 
The gravitational energy of the ring  resulting from its own mass
distribution is ${\cal U}$ while, ignoring any internal
degrees of freedom in ring particles,  $(\sum_{i=1}^{N} E_i) + {\cal U}$
gives the total energy content of the ring. 

\noindent The ring gravitational energy is
\be {\cal U} = -{G\over 2}
 \sum_{i=1}^{N}\sum_{j=1\ne i}^{N} { m_im_j \over |{\bf r}_i - {\bf r}_j|},\ee
where ${\bf r}_i$ and ${\bf r}_j$ are the position vectors of masses $i$ and $j$
respectively.

Further we suppose that some of the contribution to the satellite torques
arises from non eccentric
 Lindblad resonances of satellites  in circular orbit with pattern
speed $\omega_k,$ $ k =1, 2, 3,...$ For these we accordingly have (Freidman
 and Schutz 1978) ${\dot E_s} = \omega_k {\dot J_s}.$ Further we suppose, as
in section \ref{RAST}, that the rest comes from a satellite with an eccentric
Lindblad resonance with azimuthal mode number $m+1.$ As there, the pattern
speed $\Omega_{PP} =(\Omega_P + m\omega_s)/(m+1),$ with $2\pi/ \omega_s$
being the period of the satellite. For this contribution ${\dot E_s} =
\Omega_{PP} {\dot J_s}.$

Using these results in ({\ref{Frt}), we find:
\be 
{d I_r \over dt}  = - {1 \over \Omega_0} 
{d {\cal U} \over dt}+ \sum_{i=1}^{N} \left({1\over \Omega_i } - {1\over
\Omega_0 }\right) {d( E_i - E_{i0})\over dt} +\sum_k\left({\omega_k\over
\Omega_0 }-1\right) {\dot J_k} +\left({m\omega_s\over (m+1)\Omega_0
}-1\right) {\dot J_{m+1}} - {{\dot E_{dissip}}\over \Omega_0} .
\label{Krt}
\ee 
Here the subscripts $k$ and $m+1$ signify torques associated with the
satellite $k$ and the eccentric Lindblad resonance respectively. Further,
here we neglect $\Omega_P$ which is small in magnitude. We recall the
condition for an outer Lindblad resonance used above , namely: 
\be 
(m+1)(\Omega - \Omega_{PP}) = -\Omega 
\ee 
should be satisfied within the ring. This means that, provided the relative
ring thickness is significantly smaller than $1/m,$ we can replace $\Omega$
by $\Omega_0$ and have approximately $\Omega_0 = m\omega_s/(m+2).$ Then
Eq.~(\ref{Krt}) may be written: 
\be {d I_r \over dt} = - {1 \over \Omega_0}
{d  {\cal U} \over dt}
+ \sum_{i=1}^{N}
\left({1\over \Omega_i } - {1\over \Omega_0 }\right) {d( E_i - E_{i0})\over
dt} +\sum_k\left({\omega_k\over \Omega_0 }-1\right) {\dot J_k} + {\dot
J_{m+1}\over m+1} - {{\dot E_{dissip}}\over \Omega_0} .  
\label{Lrt}
\ee

It is helpful to use the fact that we are dealing with a narrow ring and
accordingly expand the first term in Eq.~(\ref{Lrt}) to second order in
$(a_i -a_0).$ Thus we obtain:
\be {d I_r \over dt} = {3\over 2\Omega_0} {d \over dt}\left( \sum_{i=1}^{N}
m_i \Omega_0^2 ( a_i - a_0)^2 -2 {\cal U}/3\right) +\sum_k\left({\omega_k\over \Omega_0
}-1\right) {\dot J_k} + {\dot J_{m+1}\over m+1} - {{\dot E_{dissip}}\over
\Omega_0}. \label{Mrt}
\ee

Eq.~(\ref{Mrt}) is the final form of the equation for the rate of
change of the total radial action for the ring. Their are several interesting
limiting forms.

\subsubsection{Circular ring subject to internal dissipation
       and satellite torques} \label{CRST}

In this limit $I_r,$ and ${\dot J_{m+1}},$ which vanish identically when the
ring eccentricity does, are zero and we have:
\be 
{3\over 2\Omega_0} {d \over dt}\left( \sum_{i=1}^{N} m_i \Omega_0^2 (
a_i - a_0)^2  -2 {\cal U}/3 \right) =-\sum_k\left({\omega_k\over \Omega_0 }-1\right) {\dot
J_k} + {{\dot E_{dissip}}\over \Omega_0} . 
\label{Mrt2}
\ee 

The importance of the gravitational energy 
term compared to the first term on the left hand side of (\ref{Mrt2})
which represents the relative kinetic energy of the ring is measured by
$ (M_R/M_*){\overline r}^2/(r_2 -r _1)^2, $ with $M_R$ being the mass of the ring,
This parameter is typically very small, being $\sim 10^{-4}$ for the $\epsilon$
ring of Uranus and so may be neglected but we shall retain it for now.

\noindent Note too that in the case of a confined ring where changes to
${\cal U}$ could only occur through changes in $e,$ this term would be even
smaller by a factor $e^2$ and thus be neglected according to our approximation scheme. 

Bearing this in mind,
eq. (\ref{Mrt2}) can be interpreted as giving an equation for the rate of spreading of
the ring. Note that the last dissipation term on the right hand side 
always causes the ring to
spread. The terms inside the summation giving the contributions of the
satellite torques are always positive definite because for 
standard circular orbit
torques angular momentum is transferred to or from the ring according as
the pattern speed exceeds or is less than the angular velocity $\Omega_0.$
That is ${\dot J_k} > 0 $ for $\omega_k > \Omega_0 $ and ${\dot J_k} < 0 $
for $\omega_k < \Omega_0.$ Thus the effect of the circular orbit Lindblad
torques is to oppose dissipation and to lead to confinement.

\subsubsection{Steady state ring with satellite torques and dissipation}\label{STSTATE}

In this case the time derivatives on the left hand side of 
 Eq.~(\ref{Mrt}) are zero and we
have:
\be {\dot J_{m+1}\over m+1} +\sum_k\left({\omega_k\over \Omega_0 }-1\right)
{\dot J_k} = {{\dot E_{dissip}}\over \Omega_0}. \label{Nrt}\ee
This expresses the balance between the satellite torques and energy
dissipation. When there are no circular orbit Lindblad resonances and
dissipation is induced entirely by the eccentric $m=1$ mode we obtain a
result identical to that given by Eq.~(\ref{compact}) in
section~(\ref{RAST}), namely: 
\be {\dot J_{m+1}\over m+1} = {{\dot E_{dissip}}\over \Omega_0}.
\label{Ort}
\ee
Note too, that as in section~{\ref{RAST}} we could add the effect of another
eccentric Lindblad resonance with azimuthal mode number $m-1.$

In this case dissipation induced by the $m=1$ mode is balanced by the
effects of an eccentric Lindblad resonance. Such a balance should exist
independently of the existence of additional circular orbit Lindblad torques
and background dissipation as the discussion in section~{\ref{RA}}
indicates.  This is because it is only the eccentric resonances that can
drive an eccentric $m=1$ mode and balance the effects of dissipation induced
by such a mode.

The way in which such resonances pump eccentricity has been illustrated in
section~\ref{RA} and it has been also discussed in the context of
accretion disks in Cataclysmic binaries by Lubow (1989) and protostellar
disks by Papaloizou, Nelson  and Masset (2001) and Papaloizou (2002). An
illustration of the interactions occurring in an eccentric ring is given in
Figure~1. A short account summarizing the interaction is as
follows:
\noi A potential perturbation acting on the ring with azimuthal mode
number $m$ arising from a satellite in circular orbit has pattern speed
$\omega_s.$ This couples to the global $m=1$ mode producing a resonant
forcing potential with azimuthal mode number $m+1.$ The pattern speed will
be $m\omega_s/(m+1).$ This excites a resonant response at an eccentric outer
Lindblad resonance. Energy and angular momentum are transferred to the ring
through wave excitation. The transfer ratio is the pattern speed
$m\omega_s/(m+1).$ However, the perturbing satellite can only deliver energy
and angular momentum in the ratio of its pattern speed $\omega_s.$ This
means that excess energy is available to be transferred into amplifying the
$m=1$ mode which is a global disturbance rather than a resonantly excited
short wavelength wave.

\begin{figure}[ht]
\begin{center}
\includegraphics[width=10.5cm,height=9.5cm]{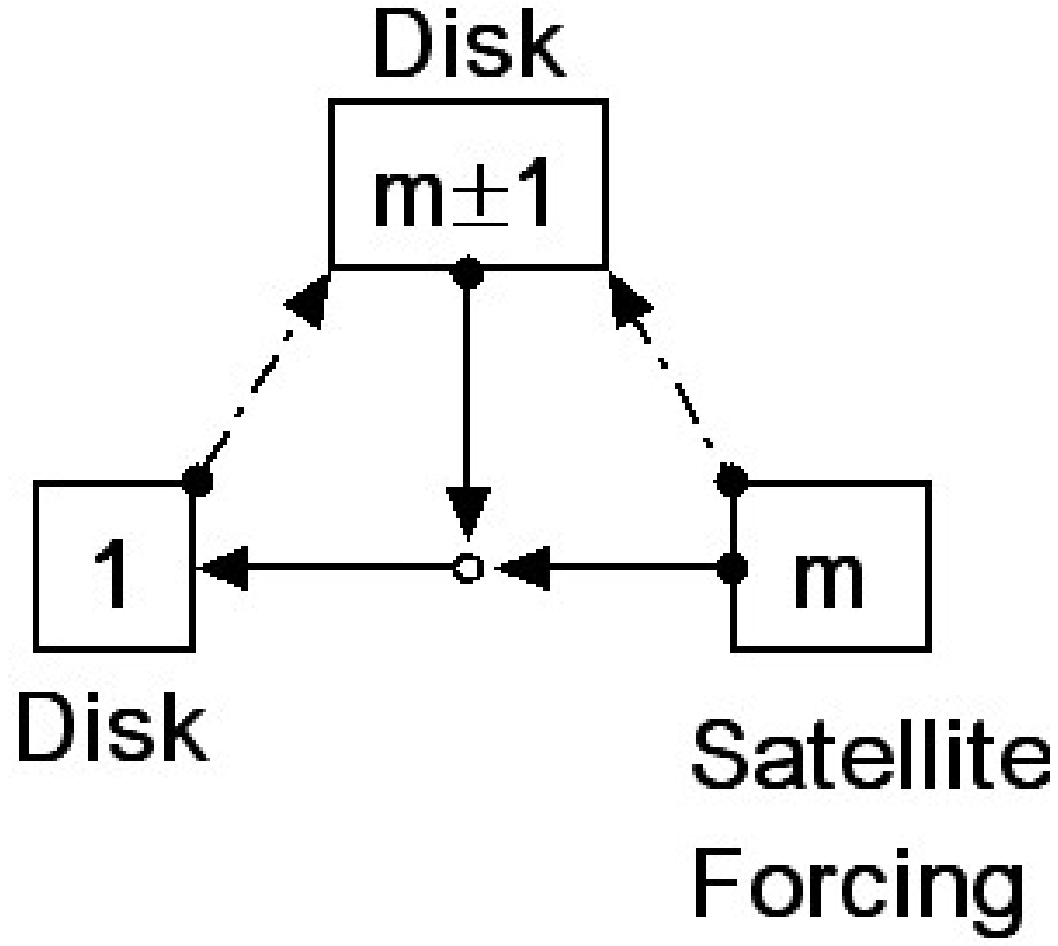}
\caption{ {\it Illustration of the interactions in an eccentric ring:} A
potential perturbation with azimuthal mode number $m$ arising from a
satellite in circular orbit is applied to the ring which is sustaining a
global $m=1$ mode manifest as a global eccentricity. The interaction between
the modes produces a further forcing perturbation with wave-number $m \pm 1$
with a possible resonant response. This response will itself interact with
the $m$-satellite forcing potential leading to a contribution to the $m=1$
mode and a positive feedback. For the maintenance of the eccentricity, the
internal dissipation produced in the ring by the $m=1$ mode, possibly
occurring mainly at the {\it pinch}, must be compensated by the input
associated with the resonant $m \pm 1$ response.}
\end{center}
\label{diag}
\end{figure}

\subsection{Application to the $\epsilon$-ring around Uranus}\label{Uranus}

We intend to show that the mechanism proposed above for the
maintenance of the eccentricity in a narrow-eccentric ring can be
applied to a real system.  We have chosen the case of the
$\epsilon$-ring of Uranus because its orbital and physical parameters
are relatively well-determined and it has known shepherd satellites.

Our goal is to verify that a reasonable
estimate for the energy dissipated in the ring is of the order
of the energy available from satellite torques (Eq.'s~\ref{compact}
and~\ref{Ort}). Here we present a simple approach, whereas a more detailed
analysis will be presented in a forthcoming paper. 

We will consider only the $47:49$ second order resonance with Cordelia,
which is the only resonance between the satellite orbit, when its eccentricity is neglected
 and the
eccentric ring (Porco and Goldreich  1987). From Goldreich and Porco (1987)
we write:
\be {\dot J_{m+1}}= 3 e^2 m^4 \ \Sigma_0\ 
\left(\frac{M_{Cordelia}}{M_{Uranus}} \right)^2 \Omega_0^2\ r_0^4  \ee 

We shall assume that the ring is highly compressed
or pinched near pericenter (Dermott  and Murray 1980) to the
extent that  the high compression results in a
 an impulsive and dissipative interaction for which
particle interactions and collisions  are important.
We comment that this situation could arise when a normal mode
grows unstably to significant amplitude such that disipation
becomes enhanced by non linear effects
leading to limitation
of the mode growth. 
The counterpart of this phenomenon  in a fluid would be the
formation of a shock.
Chiang and Goldreich (2000) and
Mosqueira and Estrada (2002) have considered models of the $\alpha$ and $\beta$ ring
in which collisional efects are important in maintaining
apse alignment against the effects of differential precession
induced by the planetary quadrupole moment and self-gravity. Mosqueira and Estrada (2002)
explicitly consider the possibility of a pinch of the type we consider
that occurs over a narrow range of azimuths near pericenter. They propose
that the density reaches a limiting value there and estimate the particle dispersion
velocity in order to calculate the effect of the pressure.
The physical state of the ring may be complex and involve some
vertical expansion ( Mosqueira and Estrada 2002). However, there may still remain
a strong dissipative interaction in the pinch.
Here we consider
the energy balance.

The origin of the
strong interaction is due to a combination of differential precession
and self-gravity (cf Eq. (\ref{gext}).)
To estimate the energy dissipated per unit mass through this strong
interaction, we adopt an 
estimate of the relative radial kinetic energy produced by the differential
precession of two orbits over one orbital period.
In making this estimate we are assuming that differential precession produced by the
planetary oblateness
is a significant factor in the balance between that, self-gravity and collisional effects.
But note that for other rings, such as the $\alpha$  ring
the balance may be between differential precession
induced by self-gravity and collisional effects with both 
of these dominating the differential precession induced by
planetary oblateness by a large factor(Chiang and Goldreich 2000, 
Mosqueira and Estrada 2002). In such cases, the potential collisional dissipation 
could be significantly underestimated by our procedure and
instead  the dominant differential precession rate
induced by self-gravity should be adopted.

For a differential precession frequency $\Delta(\omega_{prec})$
the pericenter of two orbits will be an angle $ {\cal P} = 2\pi \Delta(\omega_{prec})/\Omega$
apart after one orbit. The relative radial velocity  associated with epicyclic
motion for this phase is $\sim  er \Omega {\cal P}.$ Assuming that some fraction
of the associated energy
is dissipated every orbital period by the whole ring we obtain  an estimate for the
dissipation rate as
\be {\dot E_{dissip}}  = 2\ \pi \beta \Sigma_0  \Delta r\ r_0^3\ e^2 \Omega  
(\Delta \omega_{prec})^2, \ee
where $\beta$ is an unknown factor measuring the efficiency
with which this energy is dissipated, which must satisfy:
$\beta < 1$.

\noindent Using Eq.~(\ref{compact}) we can write:
\be
\label{eps1}
3e^2 \Sigma_0 \left( \frac{m^4}{m+1} \right) 
\left(\frac{M_{Cordelia}}{M_{Uranus}} \right)^2 \Omega_0^2\ r_0^4 =
2e^2 \Sigma_0  \beta \pi\ r_0^3\  \Delta r\ (\Delta \omega_{prec})^2
\ee

\noindent For the mean ring sem-major axis and  surface density we adopt
$r_0 = 5.11\ 10^{9} cm$ and $\Sigma_0 = 30.0\ g/cm^2$ respectively.
For the mean ring width and eccentricity 
we take $\Delta r = 5.4\ 10^{6} cm$ and $e = 0.008$ respectively.
The differential precession frequency between the outer and inner ring
edges $\Delta \omega_{prec}=  1.02\ 10^{-9} s.$ 
The mases of Uranus and Cordelia are taken to be 
$M_{Uranus} = 8.7\ 10^{28} g$ and 
$ M_{Cordelia} = 5.0\ 10^{19} g$ respectively.
The azimuthal mode number associated with the resonance is $m = 47.$

\noindent Using Eq.~(\ref{compact}) we get
for Eq.~(\ref{eps1}): $
5.75\ 10^{15} erg = \beta\ \times 9.13\ 10^{15} erg$. 
 Here, in order to make simple
estimates, we have used the same
values of $e$ and $\Sigma_0$ for both sides.
But we should remember that as distinct physical
processes operating in different ring locations are involved,
these may differ. Nonetheless  this remarkable agreement
shows that the eccentricity in the $\epsilon$-ring of Uranus can be
maintained due to the balance established between the satellite torque and
the collisional dissipation. 

\noindent However, in estimating the dissipation we only included
the effects of a pinch at pericenter. We should also consider
the possible magnitude of viscous dissipation 
arising in the general background flow.
To characterize the magnitude of the kinematic viscosity $\nu,$
we use $\nu_0~=~H^2~\Omega,$ with $\Omega= 2\times 10^{-4} s^{-1}$
and  $H= \Sigma_0/\rho_0 = 30 cm,$ being the fiducial semi-thickness
associated with a monolayer with optical depth unity,
having taken $\rho_0 = 1 g/cm^{-3}.$ 

\noindent The rate of disipation associated with background Keplerian flow
is $9\nu \Omega^2/4$ per unit mass (Lynden-Bell  and Pringle 1974).
Then 
\be {{\dot E_{dissip}}\over \Omega } = 4.5\ \pi \Sigma_0 r_0 \Delta r \nu \Omega. \ee
To compare with the above  we get ${\dot E_{dissip}}/\Omega = 4.122 \ 10^{14} ( \nu/\nu_0) \  erg.$
This is somewhat less than that 
estimated from the pinch effect but it   could be comparable
depending on the magnitude of the effective viscosity.

\noindent The background dissipation rate per unit mass
produced by the eccentricity is
estimated to be 
$\nu r_0^2 (\Delta e )^2 \Omega^2/ (\Delta r)^2$ 
(see Borderies, Goldreich  and Tremaine 1983). 
Thus this gives a rate which is factor  $4q^2/9,$   being about
 a factor of five  smaller than that acting
from the background shear. 

\noindent All of these estimates are compatible with the satellite
interaction  being adequate to resupply energy losses arising from both the
existence of the eccentric mode and  the background shear. 
\subsubsection{Timescales and small parameters} 
In view of the approximations 
made in the theoretical description it is of interest to compare the
time scales associated with the different processes involved.
The fastest time scale is the orbital timescale $\Omega^{-1} =5 \ 10^3 s,$
followed by the time scale for differential precession
$(\Delta \omega_{prec})^{-1} =  10^{9} s,$
and the time scale for the eccentricity to decay were there
to be no input from satellite torques 
$((\Delta r)^2/\nu_0) (e/\Delta e)^2 = 1.6 \ 10^{16}s,$
where we took the eccentricity difference across the ring
$\Delta e = 0.1 e.$ 

\noindent However, if as we have postulated, the 
balance is between satellite torques and dissipation in the pinch,
our discussion above indicates, because of the larger
dissipation rate,  a time scale two orders of magnitude
faster of $\sim  \ 10^{14}s,$

 \noindent These timescales 
define a small parameter $\sim 10^{-5}$
being to order of magnitude
the ratio of orbital time to  differential precession  time,
enabling justification of a slow mode approximation.
 It is also to order of magnitude the ratio of
differential precession time to  eccentricity evolution time
justifying a treatment of the $m=1$ mode neglecting
in the first instance dissipative effects and apsidal
twists. These can then be assumed to cause slow evolution of
the mode amplitude as has been done here.

\noindent  Another interesting point is that the pinch and
satellite may provide the most important dissipative effects in the ring.
This raises the possibility that this phenomenon also
dominates the confinement process. This is a complex issue
that remains to be investigated.

\newpage
%
%\begin{appendix} \label{A.1}
\section*{Appendix 1}
\section*{Forcing Potentials and Ring Eccentricity Driving}

We here consider the eccentricity excitation in a differentially rotating 
ring through the gravitational perturbation of an object in circular orbit.
The issue is which terms in the forcing potential can produce
eccentricity excitation. We consider a Lagrangian
description. Then if fluid elements
are on elliptical trajectories, an expansion of the forcing potential 
for a particular fluid element in powers of the eccentricity is possible. 

\noindent The analysis in this paper studied the excitation of eccentricity
through the ring response to the first order terms $\propto$ the ring
eccentricity $e,$  associated with a global $m=1$ mode,
in the expansion
of the satellite potential coupling back through the satellite
potential itself 
to the global $m=1$ mode of the system.

\noindent The issue arises as to whether the lowest order terms
in the satellite potential expansion that exist even when $e=0$  can
participate in ring  eccentricity excitation through 
the induced energy and angular momentum
transfer they produce. 
We here demonstrate that such an excitation does not occur. 

\subsection*{ Form of the forcing potential}
The forcing potential due to an object in circular orbit
is of the form
\be \psi_f (r, \theta, t) = \psi_f (r, \theta -\omega t).\ee
Here $\omega$ is the orbital frequency and the pattern speed of the
induced response as viewed in an inertial frame.

\noindent The energy transfer rate to the ring can be written quite generally
in the form
\be {d E \over dt} = \int \Sigma {\partial \psi_f \over \partial t} rdrd\theta, \ee
with the integral being taken over the ring.

\noindent The angular momentum transfer rate  may be similarly written
\be {d J \over dt} = -\int \Sigma {\partial \psi_f \over \partial \theta} rdrd\theta. \ee
Accordingly, because of the functional dependence
on $\theta$ and $t,$ we have the general result that
\be {d E \over dt} = \omega {d J \over dt}.\ee
The general idea of disk satellite interaction theory is that this
transfer occurs through either the excitation of a wave at a  Lindblad resonance
or directly at a corotation resonance ( Goldreich  and Tremaine 1981). In 
the situation considered here there is no corotation resonance
and the transfer occurs locally as a  wave is excited.
 At this point the wave can be regarded
as  propagating freely with the exciting potential playing no further role.
From the above it is clear that the wave is such that its energy and angular
momentum content are related by $E_{wave} = \omega J_{wave}.$
Without loss of generality we can assume that $\omega >\Omega$
so that the interaction transfers wave energy and
angular momentum {\it to} the ring.

\noindent In fact ,  energy and angular momentum transfer
to the ring does not occur until the wave dissipates (Goldreich  and Nicholson 1989). 
To examine how the transfer and  dissipation takes place we move into the frame 
rotating with angular velocity $\omega$ in which the response appears stationary.
As the forcing potential plays no further role, we can write
\be {d J_{C} \over dt} = -\epsilon_{\nu}, \ee
where $J_{C}$ is what would be the total Jacobi constant in the absence of dissipation
and the energy dissipation rate through viscosity or collisions is $\epsilon_{\nu}.$
From the point of view of the inertial
frame, the Jacobi constant per unit mass can be written, neglecting
pressure forces without loss of generality,   as
\be j_{C} = {1\over 2}\left( \left({dr\over dt}\right)^2 + r^2\left({d\theta \over dt}\right)^2\right)
 -{GM_*/r} +\psi_p - \omega r^2 \left({d\theta \over dt}\right).\ee
Accordingly $ J_{C} = E_{ring} - \omega J_{ring},$
with $E_{ring}$ and $J_{ring}$ being the energy and angular momentum content
of the ring.
Thus as the wave dissipates
\be {d E_{ring} \over dt} - \omega {d J_{ring} \over dt} =  -\epsilon_{\nu}.\ee
But from total angular momentum conservation
$ d J_{ring} / dt = -d J_{wave } / dt  $
so that
\be {d E_{ring} \over dt}= -\omega{d J_{wave} \over dt} -\epsilon_{\nu}.\ee
As  angular momentum is transferred from the wave to the ring, dissipation occurs.
This is
because when
angular momentum is transferred to local ring circular orbits with angular velocity $\Omega,$
there must at least
 be an associated  energy transfer rate  $d E_{ring}/dt = -\Omega d J_{wave } / dt.$
In that case this  means that
\be  (\Omega - \omega) {d J_{wave} \over dt} =  \epsilon_{\nu}.\ee
In fact the above expression states that as the wave is completely dissipated,
all the free energy associated with making the angular momentum transfer
to ring material on circular orbits while maintaining them on circular orbits,
which is the path that maximizes the free energy available,
is dissipated
by viscosity. There is accordingly none left to make the ring globally eccentric.

This situation is the expected one because although 
there is some free energy produced through the disk satellite interaction
that takes place through the lowest order terms in the potential expansion in powers
of $e,$  for cases of interest, where the forcing potential
has a high value of $m,$
this is created in the form of a free  wave with   the same azimuthal mode number $m$
and is  dissipated quickly and  locally as described above. In such a situation,
the associated radial motions being of high $m$ are not long lived and  do not constitute global
eccentricity. Further, in the excitation region
where the forcing potential is important, they cannot  couple to a global $m=1$  mode
with very slow pattern speed( in fact their pattern speed is preserved).
 In order to  couple to global long lived modes with slow pattern speed,
terms that are first order in $e$ are required.
These in turn produce torques and energy transfer rates which for small $e$
 are $\propto e^2$  ( see eg. Eqs. (74 - 77) of Goldreich  and Tremaine 1981).
In fact in this limit we would obtain the same torques in this work.
The thing to remember here however, is that although
produced locally,  the action is transfered
to a normal mode rather than an individual particle orbit.

%\end{appendix}

%\begin{appendix} \label{A.2}
\section*{\bf {Appendix 2}}
\section*{The Rate of Eccentricity Change and the Apsidal Twist}
In this appendix we discuss the effects of self-gravity
on the rate of change of total ring radial action and also on  ring mean 
eccentricity. Self-gravity turns out not to
enter into the expression for the time rate of change of radial action.
However, it does appear in the rate of change of mean ring eccentricity
provided there is a non zero apsidal twist, or the apsidal lines of the elliptical
streamlines within the ring are not aligned. As the rate of change of total
radial action is related to dissipation within the ring, the relation
involving apsidal twist and rate of change of mean eccentricity establishes
a connection between apsidal twist and dissipation. 

\subsection*{Self-Gravity and the Rate of Change of Radial Action}
Our starting point here is Eq.(\ref{rmot}) which governs
the radial component of the Lagrangian displacement:
\be 2\Omega {\partial^2 \xi_r\over \partial t \partial \theta_0} - \xi_r
(\Omega^2 - \kappa^2 )= f_{r} - \Delta \left( {\partial \psi' \over
\partial r} \right) + 2\Omega Q_{\theta_0}.\ee
In section \ref{RACT} we obtained an equation for the rate of change of total radial
action by multiplying Eq. (\ref{rmot}) by $\partial \xi_r / \partial \theta_0$
and integrating over the mass of the ring. We here show that  the forces arising
from self-gravity do not contribute within the approximation scheme
adopted here. In this scheme, in which the gravitational
force is approximated as equivalent to that of an infinite plane sheet
(see section \ref{SGterm} ) the  acceleration due to self-gravity is taken to be
non zero only in the radial
direction and given by    
\be
- \Delta \left({\partial \psi_{SG} \over \partial r} \right) =-  {2G\over {\overline r}} \
\int_{r_1}^{r_2} \frac{\Sigma(r_0')}{ r_0 + \xi_r - r_0' - \xi_r'}r_0' dr_0'.
\label{gss2}
\ee
Here we recall that both primed variables  being functions of $r_0',$ and unprimed variables
being functions of $r_0,$ are functions of the same angular coordinate $\theta_0.$

\noindent The potential contribution of self-gravity
 to the rate of change of total radial action
is found from Eq. (\ref{rmot}) to be ( see Eq. (\ref{consv}))
\be
{\partial \over \partial t} \left( \int \Sigma_0 \Omega \left(
{\partial\xi_r\over \partial \theta_0}\right)^2 r_0 dr_0 d\theta_0\right) = - \int
\Sigma_0 {\partial \xi_r\over \partial \theta_0} 
\Delta \left({\partial \psi_{SG} \over \partial r} \right)
 r_0 dr_0 d\theta_0 \label{gss3}.
\ee
It is a simple matter to show after inserting (\ref{gss2})
into (\ref{gss3}) that the contribution to the rate of change
of radial action is
\be
{\partial \over \partial t} \left( \int \Sigma_0 \Omega \left(
{\partial\xi_r\over \partial \theta_0}\right)^2 r_0 dr_0 d\theta_0\right) =
 - {2G\over {\overline r} }\int_{r_1}^{r_2}
\int_{r_1}^{r_2} {\partial\xi_r\over \partial \theta_0}\frac{\Sigma(r_0') \Sigma(r_0)}
{ r_0 + \xi_r - r_0' - \xi_r'} 
r_0' r_0 dr_0 dr'_0 d\theta_0 .\ee
It is straightforward by first of all noting that the integral 
on the right hand side is unchanged when primed and unprimed variables are interchanged
and so can be replaced by half the sum of the original form and the form
obtained after such an interchange. When this is done, the integrand is easily seen 
to be expressible as a derivative with respect to $\theta_0$ with the consequence
that it integrates to zero. Thus self gravity does not contribute to the rate of change
of radial action  we calculate here. This result is in line with 
 section \ref{RACTS}.

\subsection*{Self-Gravity and the Apsidal Twist}
Our starting point here is again  Eq.(\ref{rmot}). However,
we now develop an equation for the rate of change of ring
mean eccentricity rather than the mean square eccentricity
which is connected to the radial action and which has been
the theme of this paper. The equation for the time rate 
of change of the mean eccentricity relates this quantity to the apsidal
twist of the elliptical streamlines. This twist in turn is related to
the dissipation present in the ring  and is accordingly
expected to be very small. Accordingly it has been neglected so far.

\noindent We begin by recalling the form of
$m=1$ displacement used above in the form 
$\xi_r(r_0) = A(r_0)\ cos(\phi_0)$ and that $\phi_0 = \theta_0 - \Omega_Pt.$
In order to consider displacements with slowly changing eccentricity
and apsidal twist we modify this to become:
\be \xi_r(r_0) = A(r_0, t)\ cos(\phi_0 - \phi_1(r_0,t)).\label{twist} \ee
Thus the amplitude can vary slowly with time and the twist
is described by the function $\phi_1(r_0,t)$ which could
also vary slowly with time.
For example the location of pericenter is given for $A >0,$ by
$ \beta = \phi_0 -  \phi_1(r_0,t) = \pi .$ When $\phi_1$ varies with radius
a twist is generated.

\noindent In order to proceed we  insert  (\ref{twist}) into
 Eq.(\ref{rmot}) multiply by $\sin(\beta)$
and average over the angle $\theta_0$ or equivalently $\phi_0$
or $\beta.$  This procedure
yields an expression for the evolution of the ring mean
eccentricity in the form:
\be   {2\Omega {\partial  A \over \partial t }}
= -{1\over \pi} \int^{2\pi}_0 \left( f_{r}  + 2\Omega Q_{\theta_0}\right )
\sin\beta d\beta +   
{2G\over \pi \overline r} \int_{r_1}^{r_2} \int^{2\pi}_0 
\frac{\Sigma(r_0')\sin\beta}{ r_0 + A\cos\beta -
 r_0' - A'\cos \beta'}r_0' dr_0'd\beta
.\ee
Here primed quantities denote evaluation at $r_0'.$
Working to first order in the small quantity $ \Delta \phi_1 = \phi_1(r_0,t)
- \phi_1(r_0',t) =  \beta' - \beta,$
the last integral may be written as
\be - {2G\over {\pi \overline r}}\int_{r_1}^{r_2} \int^{2\pi}_0
\frac{  A' \Sigma(r_0')\Delta \phi_1\sin^2\beta}{ (r_0  -
 r_0' + (A -  A') \cos\beta)^2} 
   r_0' dr_0'd\beta
= -  {4 G\over {\overline r}}
\int_{r_1}^{r_2} \frac{ A'\Sigma(r_0')\Delta \phi_1}{(r_0  -r_0')^2} 
{I(q) \over q}  r_0' dr_0' \ee
Accordingly the evolution equation for mean eccentricity is
\be   {2\Omega {\partial  A \over \partial t }}
= -{1\over \pi} \int^{2\pi}_0 \left( f_{r}  + 2\Omega Q_{\theta_0}\right )
\sin\beta d\beta - {4 G \over {\overline r}}
\int_{r_1}^{r_2} \frac{ A'\Sigma(r_0')\Delta \phi_1}{(r_0  -r_0')^2}
{I(q) \over q}  r_0' dr_0'. \ee
Note again that the last integral again needs to be handled with caution
as it is singular and requires to be interpreted in
the principal value sense. 

\noindent One can now obtain an expression for the rate of change
of eccentricity of the entire ring by integrating over the mass.
Thus
 $$   2\Omega {\overline r} \int_{r_1}^{r_2} \Sigma {\partial  e \over \partial t } r_0 dr_0
= - {1\over \pi} \int_{r_1}^{r_2}\int^{2\pi}_0 \Sigma  \left( f_{r}  + 2\Omega Q_{\theta_0}\right )
\sin\beta d\beta r_0 dr_0 $$
 \be \hspace {3cm}   + {2 G \over {\overline r}}
\int_{r_1}^{r_2} \int_{r_1}^{r_2} \frac{ \Sigma(r_0') \Sigma(r_0)\Delta \phi_1}{(r_0  -r_0')}
I(q)   r_0' dr_0'  r_0 dr_0.  \label{FAT}\ee
Here we recall that $A= {\overline r} e$
and we have made use of the symmetry properties of the second integral
with respect to interchange of $r_0,$ and $r_0'.$

\noindent The first term on the right hand side 
 corresponds to the effects of collisions and external satellites.
We shall not dwell on this further here apart from noting that in the case
of collisions the integral can be related to the total momentum change induced by them
in different coordinate directions in the reference frame rotating with the eccentric pattern.
As collisions conserve momentum it can be argued that the net effect is small 
(cf the two streamline model of Borderies, Goldreich  and Tremaine 1983).
Thus  when external satellites are absent, there is a direct connection between
apsidal twist through the second term and the rate of mean eccentricity decay.
Equation (37) of the two stream model
of  Borderies Goldreich  and Tremaine (1983) is seen to be a discretized
form of   Eq. (\ref{FAT}).

\noindent The correct interpretation of the above is not that there
is a new dissipation mechanism, apart from collisions,
causing eccentricity decay but that the dissipative processes
occurring in collisions produce eccentricity decay as 
described through the evolution 
of the total radial action ( see Eq. (\ref{Mrt}) 
or equivalently Eq. (\ref{compact}) )
(not in fact considered by  Borderies Goldreich  and Tremaine 1983)
and also a radial dependent
phase shift in the eccentric pattern or an apsidal twist,
described by Eq. (\ref{FAT}).

\noindent  Although one must be cautious 
in using (\ref{FAT}) on account of its sensitivity
to the local  gradient of $\phi_1$ or the local apsidal twist,
we shall nonetheless make some simple estimates to compare with 
results obtained from the two streamline model of 
Borderies, Goldreich  and Tremaine (1983).
 Dropping effects due to external satellites and collisions
and simply equating characteristic 
magnitudes on the two sides of Eq. (\ref{FAT}) gives
\be
{\dot e} = { G \Sigma \Delta \phi_1 \Delta e 
\over  \Delta r \Omega} .\ee
This expression gives the same scaling relationship as Eq. (37)
of Borderies Goldreich  and Tremaine (1983).
These authors also conclude that the apsidal twist is consistent
with viscous/collisional
dissipation as  is represented in Eqs. ((\ref{Mrt}) and (\ref{compact})).
This form of dissipation  could  well be dominated by an  impulsive
interaction in a pinch at pericenter (see section \ref{Uranus} above). 

When effects due to satellites are absent, as shown by
Borderies Goldreich  and Tremaine (1983),
for a thin ring in which there is a small spatial variation of eccentricity,
the apsidal twist can be  related to the rate of energy dissipation and the
eccentricity.
We can find such a relation from Eqs. (\ref{FAT}) and (\ref{compact}) after having used
Eqs. (\ref{edeff}) and (\ref{edeff1}) to express the right hand side of the latter
in terms of the eccentricity $e,$ assuming that this does not vary with position in the ring.
One obtains from Eq. (\ref{FAT})
 \be   2\Omega {\overline r} \int_{r_1}^{r_2} \Sigma r_0 dr_0
{d e \over d t }
 = {2 G \over {\overline r}}
\int_{r_1}^{r_2} \int_{r_1}^{r_2} \frac{ \Sigma(r_0') \Sigma(r_0)\Delta \phi_1}{(r_0  -r_0')}
I(q)   r_0' dr_0'  r_0 dr_0.  \label{FAT1}\ee
One also obtains from Eq. (\ref{compact}) that
\be   2\pi \Omega {\overline r}^2 \int_{r_1}^{r_2} \Sigma r_0 dr_0
e{d e \over d t } = {-{\dot E_d} \over \Omega}. \label{FAT2}\ee
Eliminating the rate of change of eccentricity from Eqs. (\ref{FAT1}) (\ref{FAT2})
we find
\be {\dot E_d}  = -{2 \pi  G \Omega  e}
\int_{r_1}^{r_2} \int_{r_1}^{r_2} \frac{ \Sigma(r_0') \Sigma(r_0)\Delta \phi_1}{(r_0  -r_0')}
I(q)   r_0' dr_0'  r_0 dr_0,  \label{FAT3}\ee
which provides a relation between the  energy dissipation, eccentricity  and the apsidal twist.

%\end{appendix}

\section*{Acknowledgements}
\noindent The authors acknowledge support from PPARC through research grant PPA/G/02001/00486.

\section*{References}

\noi Borderies, N., Goldreich, P. and Tremaine, S.D., 1983. The dynamics
of elliptical rings. {\it Astron. J.} {\bf 88}, 1560-1568.

\noi Chiang, E.I., Goldreich, P. 2000. Apse Alignment of Narrow
Eccentric Planetary Rings.
{\it Astrophys. J.} {\bf 540}, 2, 1084-1090.

\noi Dermott, S.F., Murray, C.D., 1980. The origin of the
eccentricity gradient and the apse alignment of the $epsilon$-ring of
Uranus. {\it Icarus} {\bf 43}, 338-349.

\noi  Longaretti, P.Y., Rappaport, N., 1995.
Viscous overstabilities in dense narrow planetary rings.
 {\it Icarus} {\bf 116}, 376-596.

\noi Lynden-Bell, D., Pringle, J.E., 1974.
The evolution of viscous discs and the origin of the nebular variables.
{\bf Mon. Not. R. Astr. Soc.} {\bf 168}, 603-637.

\noi Mosqueira, I., Estrada, P.R., 2002. Apse Alignment of the Uranian
Rings. {\it Icarus}  {\bf 158}, 2, 545-556.

\noi Friedman, J. L., Schutz, B.F., 1978. Lagrangian perturbation
theory of non-relativistic fluids. {\it Astrophys. J.} {\bf 221},
937-957.

\noi Goldreich, P., Nicholson, P.D., 1989. Tides in rotating fluids.
{\it Astrophys. J.} {\bf 342},
1079-1084.

\noi Goldreich, P., Porco, C.C., 1987. Shepherding of the Uranian
rings. II. Dynamics. {\it Astron. J.} {\bf 93}, 730-737.

\noi Goldreich, P., Tremaine, S.D.,  1978.  The excitation and
evolution of density waves. {\it Astrophys. J.} {\bf 222}, 850-858.

\noi Goldreich, P., Tremaine, S.D.,  1979. Precession of the epsilon
ring of Uranus.  {\it Astron. J.} {\bf 84}, 1638-1641.

\noi Goldreich, P., Tremaine, S.D., 1981. The origin of the
eccentricities of the rings of Uranus. {\it Astrophys. J.}
{\bf 243}, 1062-1075.

\noi Goldreich, P. and S. Tremaine. 1980. Disk-satellite interactions.
{\it Astrophys. J.} {\bf 241}, 425-441.

\noi Graps, A. L., Showalter, M.R., Lissauer, J.J., Kary, D.M., 
1995. Optical Depths Profiles and Streamlines of the
Uranian (epsilon) Ring. {\it Astron. J} {\bf 109}, 2262-2273.

\noi Lynden-Bell, D. Ostriker, J.P., 1967, On the stability of
differentially rotating bodies. {\it M.N.R.A.S.} {\bf 136}, 293-310.

\noi Lubow S.H. 1989. On the dynamics of mass transfer over an
accretion disk. {\it Astrophys. J.} {\bf 340}, 1064-1069.

\noi Papaloizou J.C.B. 2002. Global m = 1 modes and migration of
protoplanetary cores in eccentric protoplanetary disks. {\it A. \& A.}
{\bf 388} 615-631.

\noi Papaloizou J.C.B. and D.N.C. Lin. 1988. On the pulsational
overstability in narrowly confined viscous rings. {\it Astrophys. J.}
{\bf 331}, 838-860.

\noi Papaloizou, J.C.B., Nelson, R.P., Masset, F. 2001. Orbital
eccentricity growth through disk-companion tidal interaction. {\it
A. \& A.} {\bf 366}, 263-275.

\noi Porco, C.C., Goldreich, P., 1987. Shepherding of the Uranian
rings. I - Kinematics. II - Dynamics. {\it Astron. J.} {\bf 93}, 724-729.

\noi  Shu F.H. 1984. Waves in Planetary rings.
In: Greenberg, R., Brahic, A.
(Eds.), Planetary Rings,  Univ. of Arizona Press, Tucson, pp. 513-561.

\noi Shu F.H., Yuan C., Lissauer, J.J. 1985. Nonlinear spiral
density waves: An inviscid theory. {\it Astrophys. J.} {\bf 291},
356-376.

\noi  Stewart, G.R., D.N.C. Lin and P. Bodenheimer. 1984.
Collision induced transport processes in planetary rings.
In: Greenberg, R., Brahic, A.
(Eds.), Planetary Rings,  Univ. of Arizona Press, Tucson, pp. 447-512

\noi Tyler, G.L.,  Eshleman, V.R., Hinson, D.P.,Marouf, E.A.,
Simpson, R.A., Sweetnam, D.N., Anderson, J.D., Campbell, J.K., Levy, G.S.,
 Lindal, G.F.  1986. Voyager 2 radio science
observations of the Uranian system Atmosphere, rings, and satellites.
{\it Science}  {\bf 233}, 79-84.

\end{document}